\documentclass[12pt,a4paper]{article}
\usepackage[utf8]{inputenc}

\usepackage{framed}
\usepackage{mdframed}

\usepackage{fullpage}
\usepackage[T1]{fontenc}

\usepackage{epsf}
\usepackage{graphicx}
\usepackage{verbatim}
\usepackage{color}	
\usepackage{bbold}
\usepackage{hyperref}
\usepackage[dvipsnames]{xcolor}

\usepackage[most]{tcolorbox}
\usepackage{empheq}
\newtcbox{\mymath}[1][]{%
    nobeforeafter, math upper, tcbox raise base,
    enhanced, colframe=blue!30!black,
    colback=blue!30, boxrule=1pt,
    #1}

\usepackage[british]{babel}
\usepackage{verbatim}
\usepackage[T1]{fontenc}
\usepackage{lmodern}
\usepackage{lipsum}
\usepackage{booktabs}
\usepackage{caption}
\usepackage{cite}
\usepackage{soul,color}
\usepackage[toc,page]{appendix}
\usepackage{float}

\usepackage{ifpdf}
\usepackage{cancel}

\usepackage{amsthm, amssymb}
\usepackage{pifont}
\usepackage{bm}
\usepackage{mathtools}
\usepackage{amstext}
\usepackage{braket}
\usepackage{multirow}
\usepackage[normalem]{ulem}
\usepackage{xcolor}
\usepackage{cancel}

\newcommand\redsout{\bgroup\markoverwith{\textcolor{red}{\rule[0.5ex]{2pt}{0.4pt}}}\ULon}

\begin{document}
\vspace{5mm}
\vspace{0.5cm}

\def\be{\begin{eqnarray}}
\def\ee{\end{eqnarray}}

\def\ba{\begin{aligned}}
\def\ea{\end{aligned}}

\def\ls{\left[}
\def\rs{\right]}
\def\lc{\left\{}
\def\rc{\right\}}

\def\p{\partial}

\def\S{\Sigma}

\def\s{\sigma}

\def\O{\Omega}

\def\a{\alpha}
\def\b{\beta}
\def\g{\gamma}

\def\ad{{\dot \alpha}}
\def\bd{{\dot \beta}}
\def\gd{{\dot \gamma}}
\newcommand{\ft}[2]{{\textstyle\frac{#1}{#2}}}
\def\ib{{\overline \imath}}
\def\jb{{\overline \jmath}}
\def\Re{\mathop{\rm Re}\nolimits}
\def\Im{\mathop{\rm Im}\nolimits}
\def\trace{\mathop{\rm Tr}\nolimits}
\def\rmi{{ i}}

\def\N{\mathcal{N}}

\newcommand{\SU}{\mathop{\rm SU}}
\newcommand{\SO}{\mathop{\rm SO}}
\newcommand{\U}{\mathop{\rm {}U}}
\newcommand{\USp}{\mathop{\rm {}USp}}
\newcommand{\OSp}{\mathop{\rm {}OSp}}
\newcommand{\Symp}{\mathop{\rm {}Sp}}
\newcommand{\Sl}{\mathop{\rm {}S}\ell }
\newcommand{\Gl}{\mathop{\rm {}G}\ell }
\newcommand{\Spin}{\mathop{\rm {}Spin}}
\newcommand*\mybluebox[1]{\colorbox{blue!20}{\hspace{1em}#1\hspace{1em}}}

\newlength{\Lnote}
\newcommand{\notte}[1]
     {\addtolength{\leftmargini}{4em}
    \settowidth{\Lnote}{\textbf{Note:~}}
    \begin{quote}
    \rule{\dimexpr\textwidth-2\leftmargini}{1pt}\\
    \mbox{}\hspace{-\Lnote}\textbf{Note:~}%
    #1\\[-0.5ex] 
    \rule{\dimexpr\textwidth-2\leftmargini}{1pt}
    \end{quote}
    \addtolength{\leftmargini}{-4em}}
\def\hc{c.c.}

\numberwithin{equation}{section}

\allowdisplaybreaks

\allowbreak



\thispagestyle{empty}
\begin{flushright}

\end{flushright}
\vspace{35pt}
\begin{center}
	    {{\fontsize{16}{20}\selectfont T-dualities and scale-separated AdS$_3$ in massless IIA on $(X_6 \times S^1)/\mathbb{Z}_2$}}

		\vspace{50pt}

    	{\fontsize{14}{20}\selectfont George~Tringas}

		\vspace{30pt}

{
{Department of Physics, Lehigh University,\\
16 Memorial Drive East, Bethlehem, PA 18018, USA}
}


\vspace{1.3cm}

E-mail: georgios.tringas@lehigh.edu

\vspace{1.3cm}

{ABSTRACT}

\end{center}

Motivated by the question of whether scale-separated AdS$_3$ flux vacua arising from G$_2$ compactifications admit an uplift to eleven-dimensional supergravity, we construct scale-separated AdS$_3$ flux vacua in massless type IIA with only O6-planes.
We first present new scale-separated solutions in massive type IIA on a G$_2$ holonomy toroidal orbifold with four smeared O6-planes, analyze their properties, and then perform a double T-duality to obtain the corresponding massless backgrounds.
In the dual frame, the internal space is locally given by a six-dimensional quotient space $X_6$ with an $\mathrm{SU}(3)$ structure of Iwasawa type times an untwisted circle $S^1$, while globally it is further modded out by a non-trivial $\mathbb{Z}_2$ quotient inherited from the G$_2$ orbifold action.
Finally, we use T-duality to derive the corresponding superpotential in massless type IIA and identify parametrically classical, scale-separated families of solutions, as well as a family with parametrically large radii, scale separation, and strong coupling, thus allowing for an uplift to eleven-dimensional supergravity.

\thispagestyle{empty} 
\setcounter{page}{0}

\baselineskip 6mm 

\newpage


{\hypersetup{hidelinks}
\tableofcontents
}
\pagestyle{empty}

\pagebreak
\pagestyle{plain}
\setcounter{page}{1}

\section{Introduction}\label{sec:introduction}

Understanding whether string compactifications can give rise to effective theories in which the cosmological constant is parametrically separated from the Kaluza--Klein scale remains an open question in string theory.
In recent years, partly motivated by the swampland program \cite{Palti:2019pca}, where the existence of scale separation has been challenged \cite{Lust:2019zwm}, the study of scale separation has become one of the most active directions in string phenomenology \cite{Coudarchet:2023mfs}.
A main goal in this context is to understand whether such vacua can be regarded as genuinely consistent string backgrounds, from the perspective of supergravity model building, holography, and the swampland. 

Following the prototypical construction of classical four-dimensional AdS vacua in massive type IIA on Calabi--Yau spaces \cite{DeWolfe:2005uu}, the so-called DGKT scenario, and its variants \cite{Camara:2005dc,Ihl:2006pp,Marchesano:2019hfb,Carrasco:2023hta,Tringas:2023vzn}, scale separation has been studied in different dimensions.
The goal has been both to clarify how broadly this property can arise in string theory and supergravity, and to identify criteria for assessing the validity of such constructions.
In particular, three-dimensional AdS models in massive type IIA have provided explicit examples in which both moduli stabilization and scale separation are realized in a parametrically classical regime \cite{Farakos:2020phe,VanHemelryck:2022ynr,Farakos:2023nms,Farakos:2023wps,Farakos:2025bwf}.
A particularly interesting feature of the three-dimensional case is that, unlike in four dimensions, scale-separated vacua have been realized not only in massive type IIA, but also in type IIB/type I supergravity \cite{Emelin:2021gzx,Arboleya:2024vnp,VanHemelryck:2025qok,Arboleya:2025lwu,Miao:2025rgf}, and even in the heterotic theory \cite{Tringas:2025bwe}.

In all the constructions mentioned above, both in three and in four dimensions, scale separation is closely tied to the presence of O-planes, whose essential role has been emphasized in \cite{Gautason:2015tig,Tringas:2025uyg}.
At the same time, these models have been subject to criticism, mainly because intersecting localized sources are often treated in the smeared approximation \cite{Banks:2006hg,McOrist:2012yc}, thereby neglecting their full backreaction.
Perturbative analyses beyond the smearing approximation \cite{Junghans:2020acz,Marchesano:2020qvg,Emelin:2022cac,Emelin:2024vug} have made it possible to incorporate backreaction effects, while attempts have also been made to understand exact localized solutions with intersecting sources directly \cite{Bardzell:2024anh}.
Additional concerns arise in massive type IIA constructions due to the presence of Romans mass \cite{Saracco:2012wc,DeLuca:2021mcj} and the associated problem of singularity resolution, although progress in this direction has been made in \cite{Junghans:2023yue}.
In this context, the existence of scale-separated solutions in massless type IIA is particularly interesting and is one of the main motivations for the present work.

Beyond the three- and four-dimensional examples, scale separation has also been explored in two dimensions \cite{Lust:2020npd,Cribiori:2024jwq}, in five dimensions \cite{Cribiori:2023ihv}, and in setups arising from six-dimensional gauged supergravity \cite{Proust:2025vmv}.
In other approaches, alternative ingredients such as Casimir energies have been considered in attempts to better understand scale separation in different frameworks \cite{Luca:2022inb,Aparici:2025kjj}.
In addition, universal features of scale separation across different dimensions have been investigated in \cite{Tringas:2025uyg}.

An important issue in the study of scale-separated vacua is to understand their consistency from a holographic point of view.
This has led to several further questions, from the understanding of the dual operators \cite{Apers:2022zjx,Apers:2022tfm}, including through explicit constructions \cite{Farakos:2025bwf}, to the more recent formulation of holographic criteria for deciding whether such constructions can admit a consistent CFT dual \cite{Bobev:2025yxp}.
From the brane picture of AdS flux vacua, these developments are part of a broader effort to clarify the holographic status of DGKT and related vacua \cite{Apers:2022vfp}.
In this context, a systematic way to reconstruct the brane configuration associated with a given AdS flux vacuum has been proposed \cite{Apers:2025pon}, and hence the form of the corresponding holographic dual.
Building on this idea, further work \cite{Bedroya:2025ltj} has challenged the consistency of these models from a holographic perspective, while more recent results \cite{Apers:2026lgi} point to a different conclusion.
From a more swampland-oriented perspective, the existence of genuine supersymmetric DGKT vacua has also been questioned \cite{Montero:2024qtz}.

The main goal of this paper is to construct parametrically strongly coupled, scale-separated vacua with large radii in massless IIA, involving only net O6-plane contributions and no non-geometric fluxes, in order to make a possible uplift to eleven-dimensional supergravity geometrically well defined.
This would allow one to assess whether scale-separated AdS$_3$ flux vacua arising from G$_2$ compactifications admit an uplift to M-theory within eleven-dimensional supergravity, in a way analogous to what was done for DGKT in \cite{Cribiori:2021djm,VanHemelryck:2024bas}.

In this paper, we begin in Section~\ref{section2} by introducing a suitable orientifold configuration in massive type IIA for which a double T-duality gives rise to massless type IIA backgrounds containing only O6-planes and no non-geometric fluxes.
In Section~\ref{section3}, we analyze the vacua of the massive theory, their stability, and the different branches of solutions that arise for different flux scalings.
In Section~\ref{section4}, we then perform the double T-duality explicitly, determine the fluxes in the massless type IIA background, and reconstruct the dual geometry.
Finally, we derive the corresponding dual superpotential and, by extremizing it, identify parametrically classical, scale-separated solutions with stabilized moduli, as well as a family of solutions that is scale separated, has large radii, and is parametrically strongly coupled, thus providing a regime for an uplift to eleven-dimensional supergravity.
In Section~\ref{section5}, we summarize our findings and discuss future directions related to the present work.

\pagebreak

\section{\texorpdfstring{G$_2$ orientifolds}{}}
\label{section2}

Scale-separated AdS$_3$ vacua have not yet been constructed in massless type IIA.
One possible strategy is to start from the known scale-separated solutions in massive type IIA \cite{Farakos:2020phe,Farakos:2025bwf} and obtain massless type IIA backgrounds by applying T-dualities.
Since our goal is to obtain massless type IIA backgrounds containing only O6-planes, we first review the G$_2$ orbifolds considered in previous constructions, explain why new configurations are needed, and then construct them.

\subsection{\texorpdfstring{G$_2$ orbifolds, shifts and fixed points}{}}

All known scale-separated AdS$_3$ solutions based on G$_2$ compactifications in massive type IIA \cite{Farakos:2020phe,VanHemelryck:2022ynr,Farakos:2023nms,Farakos:2023wps,Farakos:2025bwf}, type IIB/type I \cite{VanHemelryck:2025qok,Miao:2025rgf}, and heterotic supergravity \cite{Tringas:2025bwe} are constructed on G$_2$ toroidal orbifolds introduced by Joyce \cite{Joyce:1996a,Joyce:1996b}. The space is a seven-torus $T^7$, modded out by the orbifold group $\Gamma$, 
\begin{equation}\label{toroidalorbifold}
    X_7=
    \frac{T^7}{\Gamma}
    \cong
    \frac{T^7}{\mathbb{Z}_2 \times \mathbb{Z}_2 \times \mathbb{Z}_2}\,,
\end{equation}
generated by three elements $\Gamma=\langle \Theta_{\alpha},\Theta_{\beta},\Theta_{\gamma}\rangle$, which act on the covering space as
\begin{equation}\label{OrbifoldGenerators}
\begin{split}
\Theta_{\alpha}&: y^{i} \mapsto (-y^{1},-y^{2},-y^{3},-y^{4},y^{5},y^{6},y^{7}) \,, \\[2pt]
\Theta_{\beta}&: y^{i} \mapsto (-y^{1},\,c_1-y^{2},\,y^{3},\,y^{4},\,-y^{5},\,-y^{6},\,y^{7}) \,, \\[2pt]
\Theta_{\gamma}&: y^{i} \mapsto (c_2-y^{1},\,y^{2},\,c_3-y^{3},\,y^{4},\,-y^{5},\,y^{6},\,-y^{7})\,,
\end{split}
\end{equation}
where $y^i$ with $i=1,\dots,7.$, denote the internal space coordinates.
Here the parameters $c_i\in\{0,\tfrac12\}$ denote possible half-shifts along the torus directions.
Taking products of these generators, one obtains the remaining non-trivial elements of the orbifold group
\begin{equation}\label{OrbifoldElements}
\begin{aligned}
\Theta_{\alpha}\Theta_{\beta} : y^i & \mapsto (y^1, c_1+y^2, -y^3, -y^4, -y^5, -y^6, y^7) \,, 
\\[0.5mm]
\Theta_{\alpha}\Theta_{\gamma} : y^i & \mapsto (c_2+y^1, -y^2, c_3+y^3, -y^4, -y^5, y^6, -y^7) \,, 
\\[0.5mm]
\Theta_{\beta}\Theta_{\gamma} : y^i & \mapsto (c_2+y^1, c_1-y^2, c_3-y^3, y^4, y^5, -y^6, -y^7) \,,
\\[0.5mm]
\Theta_{\alpha}\Theta_{\beta}\Theta_{\gamma} : y^i & \mapsto (c_2-y^1, c_1+y^2, c_3+y^3, -y^4, y^5, -y^6, -y^7) \,. 
\end{aligned}
\end{equation}

We now introduce the orientifold involution relevant for our type IIA compactification. To obtain $\mathcal N=1$ supersymmetry in three dimensions, we start with a space-filling O2-plane whose $\mathbb Z_2$ target-space involution acts by reflection on all internal coordinates,
\begin{equation}\label{O2involution}
\sigma_{\text{O}2}:\ y^i\mapsto -\,y^i\,,
\qquad i=1,\dots,7\,.
\end{equation}
Composing the orientifold action $\sigma_{\text{O}2}$ with the elements of the orbifold group $\Gamma$ produces the corresponding induced involutions
\begin{equation}\label{web}
\begin{split}
\sigma_{\alpha}=\Theta_\alpha\sigma_{\text{O}2} : y^i & \mapsto (y^1, y^2, y^3, y^4, -y^5, -y^6, -y^7) \,, 
\\[2pt]
\sigma_{\beta}=\Theta_\beta\sigma_{\text{O}2} : y^i & \mapsto (y^1, c_1+y^2, -y^3, -y^4, y^5, y^6, -y^7) \,,
\\[2pt]
\sigma_{\gamma}=\Theta_\gamma\sigma_{\text{O}2} : y^i & \mapsto (c_2+y^1, -y^2, c_3+y^3, -y^4, y^5, -y^6, y^7) \,, 
\\[2pt]
\sigma_{\alpha\beta}=\Theta_{\alpha}\Theta_{\beta}\sigma_{\text{O}2} : y^i & \mapsto (-y^1, c_1-y^2, y^3, y^4, y^5, y^6, -y^7) \,, 
\\[2pt]
\sigma_{\alpha\gamma}=\Theta_{\alpha}\Theta_{\gamma}\sigma_{\text{O}2} : y^i & \mapsto (c_2-y^1, y^2, c_3-y^3, y^4, y^5, -y^6, y^7) \,, 
\\[2pt]
\sigma_{\beta\gamma}=\Theta_{\beta}\Theta_{\gamma}\sigma_{\text{O}2} : y^i & \mapsto (c_2-y^1, c_1+y^2, c_3+y^3, -y^4, -y^5, y^6, y^7) \,,
\\[2pt]
\sigma_{\alpha\beta\gamma}=\Theta_{\alpha}\Theta_{\beta}\Theta_{\gamma}\sigma_{\text{O}2} : y^i & \mapsto (c_2+y^1, c_1-y^2, c_3-y^3, y^4, -y^5, y^6, y^7) \,. 
\end{split}
\end{equation}
As is clear from these expressions, the choice of the shift parameters $c_i$ determines which involutions have fixed loci and thus can be interpreted as O6-planes. We therefore distinguish below between the different cases:
\begin{itemize}
    \item In the orbifold constructions of \cite{Farakos:2020phe,Farakos:2023nms,Farakos:2023wps}, the generators in \eqref{OrbifoldGenerators}, and thus the involutions in \eqref{web}, are chosen with $c_1=c_2=c_3=0$, and produce seven sets of O6-planes. This yields the ``maximal'' set of O6-plane configurations
    \begin{equation}
    \{{\mathrm{O6}}\}_7=\{{\mathrm{O6}_{\alpha},\mathrm{O6}_{\beta},\mathrm{O6}_{\gamma},\mathrm{O6}_{\alpha\beta},\mathrm{O6}_{\alpha\gamma},\mathrm{O6}_{\beta\gamma},\mathrm{O6}_{\alpha\beta\gamma}}\} \,.
    \end{equation}
    In this case, the net O2-plane contribution to the scalar potential is not required for moduli stabilization and can be canceled against D2-branes \cite{Farakos:2020phe}.
    However, performing double T-dualities on such maximal configurations leads to massless type IIA backgrounds containing O4-planes in addition to O6-planes.
    Moreover, since all three-cycles are threaded by NSNS flux, non-geometric fluxes are generated in the dual theory.
    As a result, the corresponding uplift to M-theory does not yield a smooth geometric background.
    Finally, these orbifolds are singular in the sense that they do not admit a resolution; see \cite{Farakos:2025bwf,Miao:2025rgf} for further discussion.
    \item 
In the orbifold used in \cite{Farakos:2025bwf,Miao:2025rgf}, the generators in \eqref{OrbifoldGenerators} are chosen so that all shifts are non-vanishing.
Due to the reduced fixed points, the set of O6-planes is then 
\begin{equation}\label{O6shifted}
\{\mathrm{O6}\}_3=\{\mathrm{O6}_{\alpha},\mathrm{O6}_{\alpha\beta},\mathrm{O6}_{\alpha\gamma}\} \,.
\end{equation}
This setup admits three possible double T-dualities, namely $\left(T_{y_1},T_{y_5}\right)$, $\left(T_{y_2},T_{y_6}\right)$, and $\left(T_{y_3},T_{y_7}\right)$, see Fig.~\ref{NewO6planesetup}, which map O6-planes to O6-planes in massless type IIA and send the NSNS flux to geometric flux.
However, the background still contains a net O2-plane contribution, which after double T-duality becomes a net O4-plane.
Unlike in the maximal O6 case, canceling the O2-planes against D2-branes and satisfying the tadpole by adjusting the flux signs does not lead to moduli stabilization.

\item 
In the ``minimal'' construction of \cite{VanHemelryck:2025qok,Tringas:2025bwe}, the generators in \eqref{OrbifoldGenerators} are chosen so that all shifts are vanishing, except for the generator $\Theta_{\gamma}$, chosen to act with a different shift along $y^5$, namely $y^5\mapsto \tfrac12-y^5$.
These setups are minimal because, in the type I/IIB background, they involve only a single symmetric metric flux and just two O5-planes, and moreover admit a resolution \cite{joyce2000compact}.
However, when one performs a double T-duality to obtain the massless type IIA description, one finds that, in the presence of the NSNS flux and the O5-planes required for moduli stabilization, avoiding non-geometric fluxes requires orientifold planes beyond O6-planes.
\end{itemize}

\subsection{Orbifold for new massive and massless IIA}\label{NEWORBIFOLD}

Motivated by the O6-plane configuration \eqref{O6shifted}, we look for an alternative contribution that can replace the O2-plane term in the scalar potential, which is required for moduli stabilization, but which does not induce a net O4-plane charge after the double T-duality.

One can generate an additional O6-plane by removing one of the shifts in \eqref{OrbifoldGenerators}.
In particular, setting $c_2=0$ makes the previously free orbifold element $\Theta_{\beta\gamma}$ develop fixed points.
As a result, in the orientifold sector the involution $\sigma_{\alpha\beta\gamma}$ develops fixed loci, giving rise to an additional O6-plane and the resulting configuration is
\begin{equation}\label{systemT}
\{\mathrm{O6}\}_{4}
=
\{\mathrm{O6}_{\alpha},\mathrm{O6}_{\alpha\beta},\mathrm{O6}_{\alpha\gamma},\mathrm{O6}_{\alpha\beta\gamma}\} \,.
\end{equation}
Then, we perform the relevant double T-duality along the coordinates $y_1,y_5$
\begin{equation}\label{eq:Tduality15}
\left(T_{y_1},T_{y_5}\right)\,,
\end{equation}
and identify the orientifold planes in the massive and massless type IIA theory
\begin{figure}[H]
\centering
\scalebox{0.91}{$
\begin{pmatrix} 
&\hspace{-0.25cm}{\rm O}2: & - & - & - & - & - & - & -  \\
&\hspace{-0.25cm}{\rm O}6_{\alpha}: & \times & \times & \times & \times & - & - & -  \\
&\hspace{-0.25cm}{\rm O}6_{\alpha\beta}: & - & - & \times & \times & \times & \times & -  \\ 
&\hspace{-0.25cm}{\rm O}6_{\gamma\alpha} : &- & \times & - & \times & \times & - & \times  \\ 
&\hspace{-0.25cm}{\rm O}6_{\alpha\beta\gamma}: &\times & - & - & \times & - & \times & \times 
\end{pmatrix}
\rightarrow
\begin{pmatrix} 
&\hspace{-0.25cm}{\rm\text{O}}4: & \times & - & - & - & \times & - & -  \\
&\hspace{-0.25cm}{\rm\widetilde{\text{O}}}6_{\alpha}: & - & \times & \times & \times & \times & - & -  \\
&\hspace{-0.25cm}{\rm\widetilde{\text{O}}}6_{\alpha\beta}: & \times & - & \times & \times & - & \times & -  \\ 
&\hspace{-0.25cm}{\rm\widetilde{\text{O}}}6_{\gamma\alpha} : & \times & \times & - & \times & - & - & \times  \\ 
&\hspace{-0.25cm}{\rm\widetilde{\text{O}}}6_{\alpha\beta\gamma}: & - & - & - & \times & \times & \times & \times 
\end{pmatrix}
$}
\caption{\textit{Left}: The orientifold involutions in the massive type IIA setup. Right: The orientifold involutions generated by the double T-duality in \eqref{systemT}. Wrapped directions are marked by $\times$, while localized positions are indicated by $-$. We emphasize once again that the O2- and O4-planes are canceled by D-branes.}\label{NewO6planesetup}
\end{figure}
As shown in Fig.~\ref{NewO6planesetup}, the coordinate $y^4$ is wrapped by all four O6-planes in both the massive and massless type IIA descriptions, whereas only the O2-plane in the former case and the O4-plane in the latter have fixed points.
As will become clear in a later section, this direction is geometrically distinguished as in the dual background the corresponding one-form $\eta^4$ is not only closed but also not related to the metric fluxes.
It is therefore natural to describe the internal space in terms of a six-dimensional manifold $X_6$, together with an additional circle direction $S^1$ associated with $y^4$.

Let us now discuss the construction from the orbifold perspective, which is equivalent to the O-plane description, since the composition of the O-plane involutions reproduces the orbifold group. From this perspective, however, the geometric structure becomes more transparent.

As follows from \eqref{OrbifoldGenerators}, the generator $\Theta_\alpha$ plays a special role, since it is the only generator in the O6-plane set \eqref{systemT} that acts as a reflection on the coordinate $y^4$.
Thus, the orbifold elements underlying the four O6-plane involutions take the form
\begin{equation}\label{group2}
\{\Theta_{\alpha},\Theta_{\alpha\beta},\Theta_{\gamma\alpha},\Theta_{\alpha\beta\gamma}\}
=
\Theta_{\alpha}\cdot
\{1,\Theta_{\beta},\Theta_{\gamma},\Theta_{\beta\gamma}\}\,.
\end{equation}
From \eqref{OrbifoldGenerators} it follows that both $\Theta_\beta$ and $\Theta_\gamma$ leave the coordinate $y^4$ invariant.
Together with their product they generate the subgroup
\begin{equation}\label{invo2}
\langle \Theta_{\beta},\Theta_{\gamma}\rangle
=
\{1,\Theta_{\beta},\Theta_{\gamma},\Theta_{\beta\gamma}\}
\cong \mathbb{Z}_2\times \mathbb{Z}_2\,,
\end{equation}
which acts non-trivially only on the six-dimensional torus parameterized by the coordinates $y^i$, with $i=1,\dots,7$ and $i\neq 4$. The corresponding six-dimensional internal space is therefore the orbifold quotient
\begin{equation}\label{q2}
X_6=\frac{T^6}{\langle\Theta_{\beta},\Theta_{\gamma}\rangle}
\cong \frac{T^6}{\mathbb{Z}_2\times\mathbb{Z}_2}\,.
\end{equation}
Accordingly, since all the generators commute, one can quotient the seven-dimensional space in stages so it can be written as
\begin{equation}\label{q1}
X_7
=
\frac{T^7}{\Gamma}
\cong
\frac{X_6\times S^1}{\widehat{\mathbb Z}_2}
=
\frac{X_6\times S^1}{\langle\Theta_{\alpha}\rangle}\,,
\end{equation}
while the circle factor is associated with $y^4$.
The remaining global $\mathbb{Z}_2$, generated by $\Theta_\alpha$ and denoted in what follows by $\widehat{\mathbb{Z}}_2$, acts non-trivially both on the circle direction and on the six-dimensional space $X_6$.

In the massive type IIA setup, where the geometry is a G$_2$ holonomy orbifold, the six-dimensional subspace may be a Calabi--Yau orbifold.
A closely related perspective was discussed in \cite{Kachru:2001je} in the context of the Joyce orbifold \eqref{OrbifoldGenerators} with all half-shifts non-zero.
In that case, the same G$_2$ orbifold admits different type IIA orientifold limits, depending on whether the M-theory circle is non-trivially acted on by one of the global generators, as in our construction.
For instance, choosing the circle along $y^7$ or $y^4$ leads to orientifold descriptions on the same Calabi--Yau target space, while choosing it along $y^5$ yields a topologically distinct one.
In this sense, the present construction is closest in spirit to one in which one isolates a special circle direction of the standard Joyce orbifold.

A related class of quotients of the form \eqref{q1} is defined by a global involution that acts anti-holomorphically on the six-dimensional Calabi--Yau space, reflects the circle, and most importantly acts freely.
In this case, the quotient is smooth and defines a barely G$_2$ manifold as discussed in \cite{Harvey:1999as,Kachru:2001je,Grigorian:2009nx}.
In our case, however, the global action is not free so the quotient is singular.
More generally, non-free quotients of this type can still be considered, as also discussed in the references above, but the existence of a smooth G$_2$ resolution is not known in full generality, see \cite{Joyce:1996b,joyce2000compact}.
Configurations with such quotients have been studied in \cite{Braun:2023zcm}, where the worldsheet CFTs of G$_2$ orbifolds obtained as quotients of Calabi--Yau times a circle are analyzed.

In the dual massless type IIA frame, the seven-dimensional geometry is described by a quotient space whose local structure consists of a six-dimensional group manifold together with an untwisted circle direction $y^4$.
This type of geometry is closely related to the description of G$_2$ structures in terms of SU(3) structures, particularly in the context of M-theory/type IIA reductions.
Constructions with this type of geometry may be found, for example, in \cite{DallAgata:2003txk,DallAgata:2005zlf,Danielsson:2014ria,Andriolo:2018yrz,delaOssa:2021cgd}, while in our case the relation is purely geometric and understood in a broader sense.


\section{\texorpdfstring{AdS$_3$ from massive IIA with 4 O6-planes}{}}
\label{section3}

In this section, we present the massive type IIA compactification on the relevant G$_2$ space and introduce the new flux ansatz associated with the new O6-plane configuration. We show how the tadpole is satisfied, and then extremize the superpotential and analyze the families of vacua characterized by the behavior of the string coupling, the internal radii, and scale separation.

\subsection{\texorpdfstring{G$_2$ holonomy and toroidal orbifold}{}}

A G$_2$ structure on the seven-dimensional internal space is specified by an associative three-form $\Phi$ and its Hodge-dual coassociative four-form $\Psi$, which in our conventions take the form
\begin{align}\label{phipsi}
    \Phi&=e^{127}-e^{347}-e^{567}+e^{136}-e^{235}+e^{145}+e^{246} \,, \\
    \Psi&=e^{3456}-e^{1256}-e^{1234}+e^{2457}-e^{1467}+e^{2367}+e^{1357} \,.
\end{align}
Here $e^{127}\equiv e^1\wedge e^2\wedge e^7$, and similarly for the other terms, with $e^i$ denoting the orthonormal frame one-forms on the internal manifold, $i=1,\dots,7\,.$.

The deformations of the G$_2$ background are encoded in seven real scalar fields $s^i\equiv s^i(x^\mu)$, where $x^\mu$ are coordinates of the external spacetime. These moduli parameterize the expansion of the associative three-form as
\begin{equation}\label{sdef}
    \Phi=\sum_{i=1}^7 s^i \Phi_i\,.
\end{equation}
A convenient basis of invariant three-forms $\Phi_i$ and dual four-forms $\Psi_i$ is given by
\begin{align}
    \Phi_i&=\left(dy^{127},-dy^{347},-dy^{567},dy^{136},-dy^{235},dy^{145},dy^{246}\right)\,,\label{basis} \\
    \Psi_i&=\left(dy^{3456},-dy^{1256},-dy^{1234},dy^{2457},-dy^{1467},dy^{2367},dy^{1357}\right)\,,\label{basis2}
\end{align}
where $y^i$ with $i=1\,,\dots,7.$, the internal space coordinates, $dy^{127}\equiv dy^1\wedge dy^2\wedge dy^7$, and similarly for the remaining basis elements, in particular, $\Phi_1=dy^{127}$ and $\Phi_2=-dy^{347}$. The three-forms in \eqref{basis}, the four-forms in \eqref{basis2}, and the top form $dy^{1234567}$ are the only invariant forms on the toroidal orbifold defined by \eqref{OrbifoldGenerators}.
The three-form basis $\Phi_i$ and the dual four-form basis $\Psi_j$ are chosen to satisfy
\begin{equation}\label{eq:PhiiPsij}
    \int_{X_7} \Phi_i\wedge\Psi_j=\delta_{ij}\,.
\end{equation}
Denoting the internal space by $X_7$, its volume is
\begin{equation}\label{volumeG2}
    \mathrm{vol}(X_7)
    =
    \frac{1}{7}\int_{X_7} \Phi\wedge\star_7\Phi
    =
    \left(\prod_{i=1}^7 s^i\right)^{1/3}\,,
\end{equation}
and we normalize the invariant top form such that 
\begin{equation}
    \int_{X_7} dy^{1234567}=1 \,.
\end{equation}
With these conventions, the Hodge duals take the form
\begin{equation}
    \star_7\Phi=\sum_i\frac{\mathrm{vol}(X_7)}{s^i}\Psi_i\,,
    \qquad
    \star_7\Phi_i=\frac{\mathrm{vol}(X_7)}{(s^i)^2}\Psi_i\,.
\end{equation}

In the massive type IIA construction, we choose the background to have G$_2$ holonomy, so that the associated G$_2$ structure is torsion-free and the internal space Ricci-flat. The fundamental forms then satisfy
\begin{equation}\label{G2closed}
    d\Phi = 0\,,
    \qquad
    d\star_7\Phi = 0\,.
\end{equation}
We now specialize to a toroidal orbifold, and the metric then reads
\begin{equation}\label{metricflat}
    ds_7^2=\sum_{i=1}^7 r_i^2\,dy_i^2 \,,
\end{equation}
with vielbeins $e^i=r_i\,dy^i$, where $r_i$, for $i=1,\dots,7$, denote the torus radii in Einstein frame.
Since we will later use the relation between the three-cycle moduli, the radii, and the overall volume, we present it here explicitly. 
Using \eqref{phipsi} and \eqref{sdef} one finds
\begin{equation}\label{eq:stor}
\begin{split}
    s^1&=r_1r_2r_7 \,,\qquad
    s^2=r_3r_4r_7 \,,\qquad
    s^3=r_5r_6r_7 \,,\qquad
    s^4=r_1r_3r_6 \,,\\
    s^5&=r_2r_3r_5 \,,\qquad
    s^6=r_1r_4r_5 \,,\qquad
    s^7=r_2r_4r_6 \,,\qquad
    \mathrm{vol}(X_7)=\prod_{i=1}^7 r_i \,.
\end{split}
\end{equation}


\subsection{Fluxes and tadpole cancellation}\label{newmassiveIIA}

The flux ansatz for the new setup is determined by the O6-plane configuration in subsection \ref{NEWORBIFOLD}, since our aim is to ensure that, after the relevant double T-dualities, the dual massless type IIA background contains only O6-planes, with no additional orientifolds and no non-geometric fluxes generated.
We therefore choose
\begin{equation}\label{F4nonminimal0}
\begin{split}
H_3 &= h\left(\Phi_1+\Phi_3+\Phi_4+\Phi_5\right)\,, \\
F_4 &= G\left(\Psi_1+\Psi_3-\Psi_4-\Psi_5\right)
-\left(N\Psi_2+M\Psi_6+R\Psi_7\right)\,, \\
F_0 &= m_0\, .
\end{split}
\end{equation}
As will become clear from the analysis, the fluxes $N$, $M$, and $R$ mainly induce anisotropies among the radii.
At the same time, keeping them distinct is essential for the emergence of the strong-coupling solution in the massless theory.
We therefore keep this generality in order to present the most general form of the solution and isolate the solution of interest.

For the O6-plane configuration in subsection \ref{NEWORBIFOLD}, the corresponding source currents can be written in the G$_2$ basis as
\begin{equation}\label{O6planesmassiveIIAcurrents}
    j_{\alpha}=\Phi_3=-dy^{567}\,,\quad\,\,\,
    j_{\alpha\beta}=\Phi_1=dy^{127}\,,\quad\,\,\,
    j_{\gamma\alpha}=\Phi_4=dy^{136}\,,\quad\,\,\,
    j_{\alpha\beta\gamma}=\Phi_5=-dy^{235}\,.
\end{equation}
We now turn to tadpole cancellation, which follows from the Bianchi identity
\begin{equation}
dF_2 = F_0 H_3 + \mu_{\mathrm{O}6}\sum_i j_i \,.
\end{equation}
Using the flux configuration in \eqref{F4nonminimal0} one finds
\begin{equation}
    0=\int_{\Sigma_{3,i}}\left(F_0 H_3+\mu_{\mathrm{O}6} j_i\right),
    \qquad
    i=\alpha,\alpha\beta,\gamma\alpha,\alpha\beta\gamma\,. \,,
\end{equation}
where $\Sigma_{3,i}$ denotes the three-cycle dual to the O6-plane current $j_i$.
Hence, the tadpole conditions constrain the combination of $H_3$ and the Romans mass $F_0$.

We now turn to the O2-plane tadpole, which is more subtle and plays a central role in our construction, and follows from the Bianchi identity
\begin{equation}
    dF_4 = F_4 \wedge H_3 + \mu_{\mathrm{O}2/\mathrm{D}2} j_7 \,.
\end{equation}
It is convenient to decompose the four-form flux into two sectors,
\begin{equation}
F_4 = F_4^G + F_4^{NMR}\,,
\end{equation}
where
\begin{equation}
F_4^G = G\left(\Psi_1+\Psi_3-\Psi_4-\Psi_5\right),
\qquad
F_4^{NMR}= -\left(N\Psi_2+M\Psi_6+R\Psi_7\right).
\end{equation}
The contribution of $F_4^{NMR}$ to the Bianchi identity vanishes identically because each term in $F_4^{NMR}$ shares at least one index with every non-vanishing component of $H_3$, therefore
\begin{equation}
    F_4^{NMR}\wedge H_3=0\,.
\end{equation}
The same holds for the $F_4^G$ sector, where with the signs in \eqref{F4nonminimal0} chosen as above, one finds
\begin{equation}\label{tadpoleF4G}
\begin{split}
    F_4^G\wedge H_3
    &=
    hG\left(\Phi_1\wedge\Psi_1
    +\Phi_3\wedge\Psi_3
    -\Phi_4\wedge\Psi_4
    -\Phi_5\wedge\Psi_5\right) \\
    &=0\,,
\end{split}
\end{equation}
and so neither of the $F_4$ components is constrained by the tadpole.

There remains the net O2-plane charge. In the present construction, since all flux contributions are arranged to cancel out, this charge must be canceled directly by adding D2-branes. Using $N_{\mathrm{O}2}=2^7$ as in \cite{Farakos:2020phe}, one finds that the cancellation condition is
\begin{equation}\label{tadpoleIIAexplicit0}
    0=\int_{X_7}(\mu_{\mathrm{O}2}+\mu_{\mathrm{D}2})\,j_7
    \qquad\rightarrow\qquad
    0=-(16-N_{\mathrm{D}2})\,,
\end{equation}
and hence $N_{\mathrm{D}2}=16$\,.
Therefore all tadpole conditions are satisfied, the net O2-plane contribution is removed from the effective action, and the flux quanta in the $F_4$ sector remain unconstrained by tadpole cancellation.


\subsection{\texorpdfstring{$\mathcal{N}=1$ 3D effective theory}{}}

To analyze the properties of the vacuum, we present the three-dimensional effective theory associated with the compactification.
The effective theory can be obtained by dimensional reduction of massive type IIA supergravity using the flux configuration of the previous subsection; for completeness, we summarize the derivation in Appendix~\ref{app:AppendixII}.
For the present purposes it is convenient to work directly with the superpotential for massive type IIA compactified on G$_2$ holonomy, as constructed in \cite{Farakos:2020phe}.
In our conventions it is given by
\begin{equation}\label{superpotential}
    P=
    \frac{1}{4\,\mathrm{vol}(X_7)^2}
    \int_{X_7}
    \left(
    e^{-\frac{\phi}{2}}\,\star_7\Phi\wedge H_3
    +e^{\frac{\phi}{4}}\,\Phi\wedge F_4
    +e^{\frac{5}{4}\phi}\,\star_7 F_0
    \right)\,.
\end{equation}
The corresponding scalar potential is
\begin{equation}\label{superpotentialtopotential}
    V = G^{IJ} P_I P_J - 4P^2\,,
\end{equation}
where $G^{IJ}$ is the inverse moduli space metric \cite{Beasley:2002db,Farakos:2020phe}, the indices run over
\begin{equation}
    I,J=s^1,\dots,s^7,\phi\,.\,,
\end{equation}
and $\phi$ is the ten-dimensional dilaton.
With this identification, the scalar potential derived from the superpotential agrees with the one obtained by direct dimensional reduction.

To evaluate \eqref{superpotential}, and for later use, it is convenient to expand the fluxes in the G$_2$ basis,
\begin{equation}\label{typeIIAfluxes}
H_3 = \sum_{i=1}^7 h^i \Phi_i \,,
\qquad
F_4 = \sum_{i=1}^7 f_4^i \Psi_i \,,
\qquad
F_0 = m_0 \,.
\end{equation}
In the present setup, for the flux ansatz in \eqref{F4nonminimal0}, this means
\begin{equation}\label{fluxx}
\begin{alignedat}{4}
f_4^1&=G \,,\qquad & f_4^2&=-N \,,\qquad & f_4^3&=G \,,\qquad & f_4^4&=-G \,,\\
f_4^5&=-G \,,\qquad & f_4^6&=-M \,,\qquad & f_4^7&=-R \,,
\end{alignedat}
\end{equation}
together with
\begin{equation}
h^1=h^3=h^4=h^5=h\,.
\end{equation}
Since the basis forms $\Phi_i$ and $\Psi_i$ are closed and co-closed on the toroidal G$_2$ background, the fluxes are harmonic and satisfy
\begin{equation}
dH_3=0\,,
\qquad
dF_4=0\,.
\end{equation}
Substituting the expansion \eqref{typeIIAfluxes} into \eqref{superpotential}, and retaining only the non-vanishing flux components, one finds
\begin{equation}\label{eq:SuperpotentialExplictSi}
\begin{split}
    P
    =
    \frac{1}{4\,\mathrm{vol}(X_7)^2}
    \Bigg(
    e^{-\frac{\phi}{2}}\mathrm{vol}(X_7)\left(\frac{h^1}{s^1}+\frac{h^3}{s^3}+\frac{h^4}{s^4}+\frac{h^5}{s^5}\right)
    +e^{\frac{\phi}{4}}\sum_{i=1}^7 f_4^i s^i
    +m_0\,\mathrm{vol}(X_7)e^{\frac{5}{4}\phi}
    \Bigg)\,.
\end{split}
\end{equation}
At this stage one could already substitute the explicit values of $h^i$ and $f_4^i$ from \eqref{fluxx}. 
However, it will be useful later to keep the superpotential in this slightly more general form, since this helps avoid sign ambiguities when performing the duality and subsequently interpreting the result in the geometry of the massless theory.


\paragraph{Kaluza--Klein states}

For the three-dimensional description to be a genuine low-energy effective theory, it must decouple from the Kaluza--Klein (KK) modes associated with the extra dimensions, which in practice requires them to be heavier than the AdS scale.
A standard criterion for this decoupling, so called scale separation, is therefore
\begin{equation}\label{scaleseparationcondition}
    \frac{\langle V\rangle}{m_{\text{KK}}^2}
    \sim
    \frac{L_{\text{KK}}^2}{L_{\text{AdS}}^2}
    \ll 1 \,.
\end{equation}

To estimate the Kaluza--Klein scale, we adopt a conservative procedure.
Although on the orbifold the natural moduli are the three-cycle volumes, we compare the vacuum expectation value with the KK mass associated with each individual torus radius.
Requiring the ratio \eqref{scaleseparationcondition} to be small for every radius ensures that the effective theory is separated from the extra-dimensional modes.

In the dual massless IIA description, the internal space is a group manifold, and an explicit computation of the full KK spectrum is more involved; see \cite{Andriot:2018tmb} and \cite{VanHemelryck:2025qok}.
For this reason, we again use the internal radii as proxies for the KK scale, rather than attempting a detailed analysis of the spectrum, since this gives a conservative but reliable criterion.

To estimate the KK masses, we work on the covering torus and impose periodic identifications\footnote{Since we are interested in the parametric scaling relevant for scale separation, we do not need to keep track of the detailed structure of the quotient space.
Passing from the covering torus to the orbifold changes only the integration domain and therefore affects the KK masses only by overall numerical factors.} $y^i\sim y^i+1$.
To proceed, we expand the ten-dimensional dilaton in Fourier modes on the internal torus as
\begin{equation}
\phi(x^\mu, y^i)=\sum_{i=1}^7\sum_{n_i=0}^{\infty}\phi_{n_i,i}(x^\mu)\cos(2\pi n_i y^i)\,,
\end{equation}
and substitute this ansatz into the ten-dimensional Einstein frame action \eqref{EinsteinFrameAction}.
We then perform a dimensional reduction of the ten-dimensional action using the ansatz
\begin{equation}\label{dimrednastz}
    ds_{10}^2=\mathrm{vol}(X_7)^{-2}ds_{3}^2+\sum_{i=1}^7 r_i^2\,dy_i^2 \,,
\end{equation}
and following the derivation in \cite{Farakos:2023nms}, one finds that the masses of the first Kaluza--Klein modes $\phi_{1,i}(x^\mu)$ have the following form
\begin{equation}
    m^2_{\text{KK},i}\equiv m^2_{\phi_{1,i}}=(2\pi)^2\frac{\mathrm{vol}(X_7)^{-2}}{r_i^2}\,.
\end{equation}
Therefore,
\begin{equation}\label{scaleseparationconditionexplicit}
\frac{\langle V\rangle}{m^2_{\text{KK},i}}
\sim
r_{i}^2\mathrm{vol}(X_7)^2\langle V\rangle \,\equiv\, r_{S,i}^2e^{-4\phi}\mathrm{vol}_S(X_7)^2\langle V\rangle \,\ll 1 \quad \forall i \,. \,,
\end{equation}
where we have used that $r_i=e^{-\frac{\phi}{4}}r_{S,i}$ and $\text{vol}(X_7)=e^{-7\phi/4}\text{vol}_S(X_7)$.


\subsection{Moduli stabilization and families of solutions}\label{modulistabilization}

We now analyze the supersymmetric AdS vacua of the construction and identify families of solutions of interest.
Since the three-dimensional effective theory is determined by the superpotential \eqref{superpotential}, we obtain supersymmetric vacua by extremizing $P$ with respect to the dilaton and the seven shape G$_2$ moduli,
\begin{equation}
    \partial_{\phi}P=0\,,
    \qquad
    \partial_{s^i}P=0\,,
    \qquad
    i=1,\dots,7\,.
\end{equation}
Solving these equations, one finds that the string coupling is
\begin{equation}\label{dilatonscaling}
     g_s\equiv e^{\phi}=\frac{2^{7/4}h}{\sqrt{3}\,m_0^{1/4}}\frac{1}{(MNR)^{1/4}} \,,
\end{equation}
and ignoring the overall numerical factor, which does not affect the parametric scaling, we see that the only parametric limit allowed by flux quantization leads to weak coupling
\begin{equation}\label{weakstringcoupling2}
    MNR\gg 1\,.
\end{equation}

Next, after performing the Weyl rescaling \eqref{WeylRescaling}, the stabilized three-cycle moduli in the string frame are given by
\begin{equation}\label{cycle2346ANIS2}
\begin{split}
    s^1_S&=s^3_S=(-1+\sqrt{3})\left(\frac{2MNR}{m_0^3}\right)^{1/4} \,,
    \qquad
    s^4_S=s^5_S=(1+\sqrt{3})\left(\frac{2MNR}{m_0^3}\right)^{1/4} \,,\\
    s^2_S&=4G\left(\frac{2MR}{m_0^3N^3}\right)^{1/4} \,,
    \qquad
    s^6_S=4G\left(\frac{2NR}{m_0^3M^3}\right)^{1/4} \,,
    \qquad
    s^7_S=4G\left(\frac{2MN}{m_0^3R^3}\right)^{1/4} \,,
\end{split}
\end{equation}
and using the relations between the radii and the three-cycle moduli in \eqref{eq:stor}, we obtain the string frame radii
\begin{align}\label{stringradii}
    r_{S,1}&=r_{S,5}=\left(\frac{2NR}{m_0M}\right)^{1/4} \,,
    \qquad
    r_{S,2}=r_{S,6}=\left(\frac{2MN}{m_0R}\right)^{1/4}\,,
    \qquad
    r_{S,4}=\frac{2^{7/4}G}{(m_0MNR)^{1/4}}\,, \nonumber \\
    r_{S,3}&=\left(1+\sqrt{3}\right)\left(\frac{MR}{2m_0N}\right)^{1/4} \,,
    \qquad
    r_{S,7}=\left(-1+\sqrt{3}\right)\left(\frac{MR}{2m_0N}\right)^{1/4} \,,
\end{align}
while the vacuum expectation value is
\begin{equation}\label{vacuumexpectationvalue}
    \langle V\rangle=-4P^2
    =
    -\frac{h^6m_0^4}{27}\frac{1}{(GMNR)^2}
    \,\sim\,
    -\frac{1}{(GMNR)^2}\,.
\end{equation}
Combining \eqref{vacuumexpectationvalue} with the Kaluza--Klein masses associated with the different internal radii, and retaining only the parametric dependence, we obtain
\begin{align}
    \frac{\langle V\rangle}{m_{\text{KK},\{1,5\}}^2}
    \sim \frac{1}{M}\,, \quad\,\,\,\,\,
    \frac{\langle V\rangle}{m_{\text{KK},\{2,6\}}^2}
    \sim \frac{1}{R}\,, \quad\,\,\,\,\,
    \frac{\langle V\rangle}{m_{\text{KK},4}^2}
    \sim \frac{G^2}{MNR}\,, \quad\,\,\,\,\,
    \frac{\langle V\rangle}{m_{\text{KK},\{3,7\}}^2}
    \sim \frac{1}{N}\,.
\end{align}

\paragraph{Families of solutions.}
In what follows, we analyze the parametric regimes defined by the behavior of the string coupling, the radii, and the scale-separation condition, and determine the corresponding conditions on the fluxes.
Since the fluxes are quantized, they take integer values and are large in the regime of interest
\begin{equation}
    G,M,N,R\gg 1\,.
\end{equation}
It is worth emphasizing that the geometrically relevant moduli in the orbifold are the three-cycle volumes rather than the individual radii.
As a result, some radii may become small while the three-cycles remain parametrically large, so that the solution still lies in a classical regime.

\begin{enumerate}
    \item \textbf{Classical solution 1.}
    
    In this regime, the solution is weakly coupled, scale separated, and all string frame radii are parametrically large
    \begin{equation}
        e^{\phi}\ll 1\,,
        \qquad
        r_{S,i}\gg 1\,,
        \qquad
        \frac{\langle V\rangle}{m_{\text{KK},i}^{2}}\ll 1
        \quad
        \,,\forall i \,.
    \end{equation}
    However, unless one imposes an additional hierarchy on the $N$, $M$, and $R$ fluxes, one obtains the condition $R<\left(G^4MN\right)^{1/3}$, together with 34 further subcases.
    For this reason, in order to present an explicit solution, we assume, only in this case, the hierarchy $N>G$, $M>N$, and $R>M$, and find that classical scale-separated solutions exist parametrically for
    \begin{equation}
    G<N<G^4 \,\quad\,\,\, \& \quad\,\,\, N<M<G^2\sqrt{N} \quad\,\,\, \& \quad\,\,\, M<R<\left(G^4MN\right)^{1/3}\,.
    \end{equation}

    \item \textbf{Classical solution 2.}
    
    This regime leads to weakly coupled, scale separated vacua, with all three-cycles parametrically large, even though two of the radii become small
    \begin{equation}
        e^{\phi}\ll 1\,,
        \quad\,\,\,
        s_{S,i}\gg 1\,,
        \quad\,\,\,
        r_{S,\{1,5\}}\ll 1\,,
        \quad\,\,\,
        r_{S,\{2,3,4,6,7\}}\gg 1\,,
        \quad\,\,\,
        \frac{\langle V\rangle}{m_{\text{KK},i}^{2}}\ll 1 \,,
    \end{equation}
    and it is realized for
    \begin{equation}\label{eq:Gwindow}
    NR<M<N^3R^3\quad\,\,\,\&\quad\,\,\,\frac{M}{(MNR)^{1/4}}<G<(MNR)^{1/2}\,.
    \end{equation}
    The behavior in which all three-cycles grow parametrically while two radii shrink is not restricted to the pair $r_1,r_5$. The same behavior occurs for the pairs $r_2,r_6$ and $r_3,r_7$, provided the fluxes satisfy the corresponding analogous conditions. These are precisely the pairs exchanged by the other two double T-dualities, $(T_2,T_6)$ and $(T_3,T_7)$, which lead to the same massless type IIA setup with four O6-planes.
\end{enumerate}


\section{\texorpdfstring{AdS$_3$ from massless IIA with 4 O6-planes}{}}\label{section4}

In this section, we first perform a double T-duality to determine the fluxes in the dual frame and the corresponding tadpole cancellation.
We then study the resulting internal geometry, identify the invariant subspaces and the untwisted scalar fields, and construct the SU(3) structure on the six-dimensional space.
Finally, we use double T-duality to derive the massless type IIA superpotential from its massive type IIA counterpart and identify the corresponding families of solutions, as characterized by the behavior of the string coupling, the radii, and scale separation.

\subsection{Double T-duality}\label{DoubleTduality}

We now dualize the NSNS and RR fluxes using the standard Buscher rules following \cite{Shelton:2005cf,Wecht:2007wu}.
Writing the NSNS flux components of the massive type IIA background in the notation of \eqref{Findicesnotation}, a T-duality along the direction $y^i$ maps the corresponding $H_3$ component into metric flux according to
\begin{equation}
    T_i:\quad H_{ijk}\ \leftrightarrow\ \tau^i{}_{jk}\,.
\end{equation}
where $\tau^i{}_{jk}$ are the structure constants of the Lie algebra associated with a group manifold.
Under the double T-duality \eqref{eq:Tduality15}, the NSNS flux components in \eqref{Hdual1} are mapped to
\begin{equation}\label{metricfluxes}
    H_{567}\leftrightarrow \tau^5{}_{67}\,,
    \qquad
    H_{127}\leftrightarrow \tau^1{}_{27}\,,
    \qquad
    H_{136}\leftrightarrow \tau^1{}_{36}\,,
    \qquad
    H_{235}\leftrightarrow \tau^5{}_{23}\,,
\end{equation}
with their signs fixed by the NSNS ansatz \eqref{F4nonminimal0} together with the G$_2$ basis \eqref{basis}, so that
\begin{equation}\label{metricfluxes2}
    \tau^5{}_{67}=-h^3=-h\,,
    \qquad
    \tau^1{}_{27}=h^1=h\,,
    \qquad
    \tau^1{}_{36}=h^4=h\,,
    \qquad
    \tau^5{}_{23}=-h^5=-h\,.
\end{equation}

In the dual geometry, the one-forms are no longer closed, but instead satisfy the Maurer--Cartan equation
\begin{equation}\label{MaurerCartan}
    d\eta^i=\frac{1}{2}\tau^i{}_{jk}\,\eta^j\wedge\eta^k\,,
\end{equation}
where in our case the structure constants satisfy both the unimodularity condition and the Jacobi identity,
\begin{equation}
    \tau^{i}{}_{ji}=0\,,
    \qquad
    \tau^{l}{}_{[ij}\tau^{m}{}_{k]l}=0\,.
\end{equation}
It is important to note that the group manifold obtained after the double T-duality has a two-step nilpotent six-dimensional Lie algebra.
More precisely, the Maurer--Cartan equations take the form
\begin{equation}\label{Maurer2}
d\eta^1 = h\left(\eta^{27}+\eta^{36}\right)\,,
\qquad
d\eta^5 = -h\left(\eta^{67}+\eta^{23}\right)\,,
\qquad
d\eta^i = 0\,, \quad i \neq 1,5 \,.,
\end{equation}
and after a linear redefinition of the left-invariant basis, this algebra can be brought to the standard Iwasawa form\footnote{The Iwasawa manifold also appears in the massive dual of DGKT discussed in \cite{Cribiori:2021djm}, in the context of M-theory uplift of scale-separated AdS$_4$ vacua.
In that case, however, the internal space is six-dimensional, and the moduli can already be stabilized by $F_2$ and $F_6$ fluxes alone.
By contrast, our massless dual background contains an additional circle direction ($\sim X_6\times S^1$) and requires $F_4$ flux components with one leg along the circle which allows for stabilizing the relevant radius.
We further remark that this manifold is included in the classification of group manifolds relevant for scale-separated AdS$_3$ constructions in type IIB \cite{VanHemelryck:2025qok} (Nilmanifold $\mathfrak{n}_6$ in the classification table there).
Nevertheless, the solution considered there does not dualize to the present setup, due to their different flux configuration.}, see \cite{Caviezel:2008ik}.

We would like to emphasize the role of the O2/D2 and O4/D4 systems in the massive and massless type IIA configurations respectively.
On the massive IIA side, although the O2-plane charge is canceled by the D2-branes, so that there is no net tadpole contribution, the associated orientifold involution still acts on all internal coordinates.
This is crucial, because such an action is incompatible with metric flux in the massive IIA frame.
Under the O2 involution, the left-hand side of the Maurer--Cartan equation is odd, whereas the right-hand side, being quadratic in the one-forms, is even.
After the double T-duality, the same system is mapped to an O4/D4 configuration in the massless IIA frame. The O4-plane charge is canceled by the D4-branes, while the involution now acts as
\begin{equation}\label{O4involution}
\begin{aligned}
\sigma_{\text{O}4}:\quad
\eta^i
&\mapsto
(+\eta^1,-\eta^2,-\eta^3,-\eta^4,+\eta^5,-\eta^6,-\eta^7)\,.
\end{aligned}
\end{equation}
Using the metric fluxes in \eqref{metricfluxes} together with the Maurer--Cartan equation \eqref{MaurerCartan}, one finds that the two sides now transform consistently under $\sigma_{\text{O}4}$ as the one-forms on the left-hand side are even, and so are the corresponding two-form combinations on the right-hand side. Thus, unlike the O2 involution in the massive theory, the O4 involution is compatible with the twisted geometry of the massless dual.

Now we move to the RR sector, and the Romans mass is mapped to an $\widetilde{F}_2$ flux supported along the two dualized directions,
\begin{equation}\label{Romansdual}
\widetilde{F}_{\mathbf{2},15}=m_0\eta^{15}\,.
\end{equation}
Here and in the following, we use the notation $F_{\mathbf{p},m_1\ldots m_p}$, introduced in \eqref{Findicesnotation}, where $p$ denotes the degree of the RR flux and the indices $m_1,\ldots,m_p$ specify the internal directions along which the corresponding component is supported.

The remaining RR fluxes are mapped according to \eqref{RRmapping}, and collecting all contributions, the RR sector of the massless type IIA background is
\begin{align}
\widetilde{F}_2 &= m_0\eta^{15} + N\eta^{26} - R\eta^{37}\,, \label{F2components}\\
\widetilde{F}_4 &= G\eta^{1346} - G\eta^{2345} - G\eta^{1247} + G\eta^{4567}\,, \label{F4components}\\
\widetilde{F}_6 &= -M\eta^{123567}\,. \label{F6components}
\end{align}


\subsection{Tadpole cancellation}\label{tadpolecancellationmassless}

Now that the metric fluxes obtained by the double T-duality are known, see \eqref{metricfluxes} and \eqref{metricfluxes2}, we can evaluate the exterior derivatives of the RR fluxes and determine how the tadpoles are satisfied in the massless type IIA background.

Using the Maurer--Cartan equation \eqref{MaurerCartan}, together with the fact that the only non-vanishing metric fluxes are those in \eqref{metricfluxes2}, one finds
\begin{align}\label{dF2nonzero}
\begin{split}
    d\widetilde{F}_{\mathbf{2},15}
    &=m_0\,d(\eta^{15})\,,\\
    &=-m_0\left(\tau^5{}_{23}\eta^{123}+\tau^5{}_{67}\eta^{167}+\tau^1{}_{27}\eta^{257}+\tau^1{}_{36}\eta^{356}\right)\,,\\
    &=-m_0 h\left(-\eta^{123}-\eta^{167}+\eta^{257}+\eta^{356}\right)\,,
\end{split}
\end{align}
whereas
\begin{equation}
    d\widetilde{F}_{\mathbf{2},26}=0\,,
    \qquad
    d\widetilde{F}_{\mathbf{2},37}=0\,.
\end{equation}
Hence the only non-vanishing contribution to $d\widetilde{F}_2$ comes from the component obtained by dualizing the Romans mass \eqref{Romansdual}.
The right-hand side of \eqref{dF2nonzero} reproduces precisely the four components of $H_3$ appearing in the massive type IIA background.
Thus, in the dual massless IIA frame, $d\widetilde{F}_2$ cancels the charge of the four O6-planes, exactly as $F_0H_3$ cancels the corresponding O6-plane charge in the massive IIA frame.

The O6-plane currents in the massless type IIA are obtained by applying the double T-duality \eqref{eq:Tduality15} to the massive IIA currents \eqref{O6planesmassiveIIAcurrents},
\begin{equation}\label{O6planesmasslessIIAcurrents}
    \widetilde{j}_{\alpha}=-\eta^{167}\,,\qquad
    \widetilde{j}_{\alpha\beta}=\eta^{257}\,,\qquad
    \widetilde{j}_{\gamma\alpha}=\eta^{356}\,,\qquad
    \widetilde{j}_{\alpha\beta\gamma}=-\eta^{123}\,.
\end{equation}
We keep the same subscripts as in the massive theory in order to indicate which orientifold involution each current descends from. Thus, the Bianchi identity for $\widetilde{F}_2$ is 
\begin{equation}
d\widetilde{F}_2=\mu_{\mathrm{O}6}\sum_i \widetilde{j}_i\,,
\qquad
i=\alpha,\alpha\beta,\gamma\alpha,\alpha\beta\gamma\,.\,,
\end{equation}
and considering \eqref{dF2nonzero} the tadpole condition implies $\mu_{\mathrm{O}6}=-m_0 h$. The metric fluxes are fixed by the dualized NSNS flux and therefore cannot scale, whereas the harmonic components $\widetilde{F}_{\mathbf{2},26}$ and $\widetilde{F}_{\mathbf{2},37}$ are not constrained and can scale parametrically.

We now turn to the four-form flux, and using the Maurer--Cartan equation \eqref{MaurerCartan} together with the metric fluxes \eqref{metricfluxes2}, one finds
\begin{equation}\label{dF4}
\begin{split}
d\widetilde{F}_4
&=
d\widetilde{F}_{\mathbf{4},1346}
+d\widetilde{F}_{\mathbf{4},2345}
+d\widetilde{F}_{\mathbf{4},1247}
+d\widetilde{F}_{\mathbf{4},4567}\\
&=
G\left(
-\tau^1{}_{27}
+\tau^5{}_{67}
+\tau^1{}_{36}
-\tau^5{}_{23}
\right)\eta^{23467}\\
&=0\,.
\end{split}
\end{equation}
Thus, the flux contribution to the relevant tadpole vanishes identically and therefore the flux $G$ is not constrained by tadpole cancellation, while the O4-planes are canceled by the appropriate number of D4-branes.

Finally, it follows immediately from \eqref{F6components} and the Maurer--Cartan equation that
\begin{equation}
d\widetilde{F}_6=0\,,
\end{equation}
consequently the flux component $M$ remains unconstrained.


\subsection{Geometry of the massless IIA background}

As shown in subsection \ref{DoubleTduality}, the duality reshuffles the fluxes, and the resulting massless type IIA background contains non-trivial metric fluxes.
The corresponding nilpotent algebra is isomorphic to the Iwasawa algebra and therefore admits a left-invariant $\mathrm{SU}(3)$ structure.
The relevant geometric data are encoded in the left-invariant one-forms $\eta^i$ which now satisfy \eqref{Maurer2}.
In particular, it is clear from \eqref{metricfluxes2} that the index $4$ does not appear in any non-vanishing structure constant $\tau^i{}_{jk}$, so $\eta^4$ is completely decoupled from the twisting.
It is therefore natural to regard the local geometry as a six-dimensional space with SU(3) structure together with an untwisted circle,
\begin{equation}
ds_7^2=ds_6^2+r_4^2(\eta^4)^2\,,
\end{equation}
where locally one may identify $\eta^4=dy^4$. Globally, however, the orbifold generator $\Theta_\alpha$ acts in the dual frame as a $\widehat{\mathbb Z}_2$ identification, so the seven-dimensional geometry is described by the quotient in \eqref{q1}.


\subsubsection{\texorpdfstring{G$_2$ splitting and $\mathrm{SU}(3)$ structure on $X_6$}{G2 splitting and SU(3) structure on X6}}\label{G2splitting}

We reformulate the G$_2$ geometry in terms of an SU(3) structure by decomposing the forms $\Phi$ and $\Psi$ with respect to the untwisted direction $e^4$, thus identifying on $X_6$ the $(1,1)$-form $J$ and the corresponding $(3,0)$-form $\Omega$.
Using the standard decomposition of a G$_2$ structure with respect to a unit one-form $v$ \cite{Chiossi:2002tw,Grigorian:2008tc}, we obtain the standard relations between the G$_2$ and SU(3) structure forms
\begin{equation}\label{PsiJOmega}
\Phi=v\wedge J+\mathrm{Re}\,\Omega \,,
\qquad
\Psi=\frac12\,J\wedge J-v\wedge \mathrm{Im}\,\Omega \,,
\end{equation}
where $\Psi=\star_7\Phi$.
The relative signs in this decomposition are a matter of convention, but they must be chosen consistently\footnote{The convention adopted for the G$_2$ splitting must be compatible with the metric and orientation induced by the SU(3) structure.
We verify that for our choice of $J$, $\Omega$, and $e^4$, the relations $\Psi=\star_7\Phi$ and $\Phi\wedge\Psi=7\mathrm{vol}_7$ hold.
From \eqref{JReOmega} and \eqref{eImOmega} one finds $\star_6\,\mathrm{Re}\,\Omega=-\,\mathrm{Im}\,\Omega$.
Using \eqref{vol6}, one obtains $\mathrm{vol}_7=-\,\mathrm{vol}_6\wedge e^4=e^{1234567}$,
where $\mathrm{vol}_D$ denotes the corresponding volume form. It follows that
\begin{equation}
\star_7(e^4\wedge J)=-\,\star_6 J=\frac12\,J\wedge J \,,
\qquad
\star_7(\mathrm{Re}\,\Omega)=e^4\wedge \star_6\mathrm{Re}\,\Omega
=-\,e^4\wedge \mathrm{Im}\,\Omega \,,
\end{equation}
and therefore
\begin{equation}
\star_7\Phi=\star_7\bigl(e^4\wedge J+\mathrm{Re}\,\Omega\bigr)
=\frac12 J\wedge J-e^4\wedge \mathrm{Im}\,\Omega
=\Psi \,.
\end{equation}}. In the following, we adopt the convention with a minus sign in front of $\mathrm{Im}\,\Omega$, following \cite{Grigorian:2008tc}. With this choice, the induced orientation is aligned with the SU(3) structure conventions of \cite{Lust:2004ig}, which allows us to read off the torsion classes directly from the differential relations of the structure forms.

Separating the terms in $\Phi$ and $\Psi$ that contain $v=e^4$, and using the basis conventions in \eqref{phipsi}, one finds
\begin{align}
    \Phi
    &=e^4\wedge\left(e^{37}-e^{15}-e^{26}\right)
      +e^{127}-e^{567}+e^{136}-e^{235}\,, \\
    \Psi
    &=e^4\wedge\left(-e^{356}+e^{123}-e^{257}+e^{167}\right)
      -e^{1256}+e^{2367}+e^{1357}\,.
\end{align}
Comparing with \eqref{PsiJOmega}, we identify the K\"ahler form in the orthonormal frame as
\begin{equation}\label{JReOmega}
    J=e^{37}-e^{15}-e^{26}\,,
    \qquad
    \frac12\,J\wedge J
    =-e^{1256}+e^{2367}+e^{1357} \,,
\end{equation}
and the real and imaginary parts of the holomorphic $\Omega$ three-form,
\begin{equation}\label{eImOmega}
    \mathrm{Re}\,\Omega=e^{127}-e^{567}+e^{136}-e^{235}\,,\quad\quad
    \mathrm{Im}\,\Omega
    =-e^{123}-e^{167}+e^{257}+e^{356} \,,
\end{equation}
and the compatible form reproducing these expressions is
\begin{equation}\label{Omega}
    \Omega=(e^1-i e^5)\wedge (e^2-i e^6)\wedge (e^7-i e^3) \,.
\end{equation}
From the $\mathrm{SU}(3)$ structure forms obtained through this decomposition, it follows that the global involution generated by $\Theta_{\alpha}$ acts anti-holomorphically
\begin{equation}
    \Theta_{\alpha}^{*}\,J=-J\,,
    \qquad
    \Theta_{\alpha}^{*}\,\Omega=\overline{\Omega}\,.
\end{equation}

As expected, the forms appearing in the decomposition \eqref{PsiJOmega} are precisely those spanning the invariant even and odd subspaces $\mathcal A^p_{\pm}(X_6)$, since the decomposition is defined on the six-dimensional forms under the involution induced by $\Theta_\alpha$.
Then, the corresponding $\widehat{\mathbb Z}_2$-invariant sectors on $X_7$ are then reconstructed as
\begin{equation}
\mathcal A^p_{\mathrm{inv}}(X_7)
=
\mathcal A^p_{+}(X_6)\oplus \eta^4\wedge \mathcal A^{p-1}_{-}(X_6)\,,
\end{equation}
where the one-form along the circle direction is odd under the global involution.
For convenience, we pass to the left-invariant basis $\eta^m$ on the six-dimensional twisted space $X_6$, defined by
\begin{equation}\label{orthonormalframe}
e^a=r_a\eta^a\,,
\qquad
a=1,2,3,4,5,6,7 \,.
\end{equation}
We then decompose the left-invariant two-forms on $X_6$ into even and odd eigenspaces under $\Theta_\alpha$, and one finds
\begin{equation}\label{2formsubspace}
\mathcal A^2_+(X_6)=0\,,
\qquad
\mathcal A^2_-(X_6)=\mathrm{span}\{\eta^{15},\,\eta^{26},\,\eta^{37}\}\,,
\end{equation}
from which it follows in particular that $J$ is odd under the global involution.
The wedge products of the odd two-forms generate the corresponding even four-form subspace
\begin{equation}\label{4formsubspace}
\mathcal A^4_-(X_6)=0\,,
\qquad
\mathcal A^4_+(X_6)=\mathrm{span}\{\eta^{1526},\,\eta^{1537},\,\eta^{2637}\}\,.
\end{equation}
For three-forms, one similarly finds
\begin{equation}\label{inv3forms}
\mathcal A^3_+(X_6)
=\mathrm{span}\{\eta^{127},\,\eta^{136},\,\eta^{235},\,\eta^{567}\}\,,
\quad\,
\mathcal A^3_-(X_6)
=\mathrm{span}\{\eta^{123},\,\eta^{167},\,\eta^{257},\,\eta^{356}\}\,,
\end{equation}
showing that $\mathrm{Re}\,\Omega$ is even, $\mathrm{Im}\,\Omega$ is odd while the unique six-form $\eta^{123567}$ is odd.
The invariant eigenspaces obtained in this way are precisely those of the standard $T^6/(\mathbb Z_2\times \mathbb Z_2)$ orbifold; see for example \cite{Palti:2019pca}.


\subsubsection{Fluxes in 6+1 language and orientifold projections}

We now rewrite the RR fluxes in terms of the SU(3) structure forms on $X_6$, together with the untwisted one-form $\eta^4$. 

As follows from the double T-duality, and in particular from \eqref{F4components}, the only RR flux with one leg along the untwisted direction is $\widetilde{F}_4$. 
Its general form is
\begin{equation}\label{tildeF4}
\widetilde{F}_4
=
\eta^4\wedge\left(
f_4^1\,\eta^{136}
-f_4^3\,\eta^{235}
+f_4^4\,\eta^{127}
-f_4^5\,\eta^{567}
\right)\,,
\end{equation}
for the specific flux choice in \eqref{fluxx}, this reduces to
\begin{equation}\label{F4dual}
\widetilde{F}_4
=
G\,\eta^4\wedge\Xi_3\,,\quad\quad
\Xi_3=\left(
\eta^{136}-\eta^{235}-\eta^{127}+\eta^{567}
\right)\,.
\end{equation}
The three-form $\Xi_3$ appearing in $\widetilde{F}_4$ belongs to the subspace $\mathcal{A}^3_+(X_6)$, however, it is not proportional to $\mathrm{Re}\,\Omega$, and it is therefore most transparent to keep $\widetilde{F}_4$ written in the explicit invariant-form basis. Its closure follows from the Bianchi identity and is indeed satisfied for the flux choice in \eqref{fluxx}.

The RR two-form does not carry a leg along the untwisted direction and is naturally expanded on the common odd two-form basis \eqref{2formsubspace}. Using \eqref{F2components}, one finds
\begin{equation}\label{tildeF2}
\widetilde{F}_2
=
m_0\,\eta^{15}
-f_4^2\,\eta^{26}
+f_4^7\,\eta^{37}\,,
\end{equation}
and thus it is expanded on the same invariant two-form subspace as the K\"ahler form $J$, although it is not proportional to it up to signs.
Nevertheless as we will see, their exterior derivatives are proportional in the present background. 

Finally, using \eqref{F6components} together with the six-dimensional volume form in \eqref{vol6}, the six-form flux can be written as
\begin{equation}\label{F6def}
\widetilde{F}_6
=
f_4^6\,\eta^{123567}\,,
\end{equation}
and its exterior derivative vanishes.

\paragraph{Orientifold projections.}

We now analyze the transformation properties of the fluxes under the orientifold involutions and verify their compatibility with the geometry.

In the presence of orientifold planes, the RR fluxes $F_p$ must transform with a definite parity under the target-space involution $\sigma$,
\begin{equation}
\sigma^{*}F_{p}=\pm F_{p}\,.
\end{equation}
Following \cite{VanRiet:2011yc}, we summarize in Table~\ref{tab:orientifoldparities} the transformation properties of the fluxes relevant for the orientifolds considered in this work:
\begin{table}[H]
\centering
\renewcommand{\arraystretch}{1.25}
\begin{tabular}{|c|c|c|c|}
\hline
Fluxes & O2 & O4 & O6 \\ \hline
$F_0$ & + & -- & + \\ \hline
$F_2$ & -- & + & -- \\ \hline
$F_4$ & + & -- & + \\ \hline
$F_6$ & -- & + & -- \\ \hline
$H_3$ & -- & -- & -- \\ \hline
\end{tabular}
\caption{Parity of the fluxes under the orientifold involutions.}
\label{tab:orientifoldparities}
\end{table}

In the massive background, one verifies that the basis forms in \eqref{basis}--\eqref{basis2}, on which the $F_4$ and $H_3$ fluxes are expanded, transform under the O6-plane involutions as
\begin{equation}
    \sigma^*_i\Phi=-\Phi\,,
    \qquad
    \sigma^*_i\Psi=+\Psi\,,
    \qquad
    \text{for}\qquad i=\alpha,\alpha\beta,\alpha\gamma,\alpha\beta\gamma\,.\,,
\end{equation}
while the same transformation properties also hold under the O2-plane involution $\sigma_{\text{O}2}$, in agreement with Table~\ref{tab:orientifoldparities}.

We now turn to the O4/O6 orientifold system. One finds explicitly that the basis forms on which the fluxes in the massless type IIA background are expanded transform under the O6-plane involutions as
\begin{equation}
    \widetilde{\sigma}^*_i\omega_j=-\omega_j\,,
    \qquad
    \widetilde{\sigma}^*_i(\eta^4\wedge\Xi)=+\eta^4\wedge\Xi\,,
    \qquad
    \widetilde{\sigma}^*_i\eta^{123567}=-\eta^{123567}\,,
\end{equation}
for $i=\alpha,\alpha\beta,\alpha\gamma,\alpha\beta\gamma\,.$,
with
\begin{equation}
\omega_1=-\eta^{15}\,,
\qquad
\omega_2=-\eta^{26}\,,
\qquad
\omega_3=\eta^{37}\,.
\end{equation}
On the other hand, under the O4-plane involution one has
\begin{equation}
    \widetilde{\sigma}^*_{\mathrm{O}4}\omega_j=+\omega_j\,,
    \qquad
    \widetilde{\sigma}^*_{\mathrm{O}4}(\eta^4\wedge\Xi)=-\eta^4\wedge\Xi\,,
    \qquad
    \widetilde{\sigma}^*_{\mathrm{O}4}\eta^{123567}=+\eta^{123567}\,,
\end{equation}
Compared with the massive O2/O6 background, the fluxes in the O4/O6 system therefore split into two sectors with different orientifold parities. This is fully consistent with the geometry obtained after T-duality and is analogous to the O5/O7 systems studied in \cite{Caviezel:2009tu}, where the basis forms are likewise characterized by their transformation properties under different orientifold involutions.

This structure can also be understood directly from the orbifold perspective, through the transformation properties of the forms on $(X_6\times S^1)/\widehat{\mathbb{Z}}_2$ in the massless theory.
The composition of the O4 involution with each of the O6 involutions reproduces the corresponding element of the orbifold group $\Gamma$
\begin{equation}
    \widetilde{\sigma}_{\mathrm{O}4}\widetilde{\sigma}_{\alpha}
    =\Theta_{\alpha}\,,
    \qquad
    \widetilde{\sigma}_{\mathrm{O}4}\widetilde{\sigma}_{\alpha\beta}
    =\Theta_{\alpha\beta}\,,
    \qquad
    \widetilde{\sigma}_{\mathrm{O}4}\widetilde{\sigma}_{\alpha\gamma}
    =\Theta_{\alpha\gamma}\,,
    \qquad
    \widetilde{\sigma}_{\mathrm{O}4}\widetilde{\sigma}_{\alpha\beta\gamma}
    =\Theta_{\alpha\beta\gamma}\,.
\end{equation}
This is consistent with Fig.~\ref{NewO6planesetup} and with the orbifold group elements listed in \eqref{OrbifoldGenerators}--\eqref{OrbifoldElements}.
It also shows that, in the massless theory, the basis forms decompose into sectors with different transformation properties under the generators of $\Gamma$, in contrast with the massive theory, where the relevant three-forms and four-forms carry the same parity under the orbifold generators.


\subsubsection{Scalar degrees of freedom}

Since the expressions above are written in the orthonormal frame $e^a$, the dependence on the scalar moduli is not yet explicit. 
The metric on $X_6$ is diagonal in this basis, while the odd two-form subspace \eqref{2formsubspace} naturally singles out the three K\"ahler moduli
\begin{equation}
t_1=r_1r_5\,,
\qquad
t_2=r_2r_6\,,
\qquad
t_3=r_3r_7\,,
\end{equation}
which control the volumes of the two-planes $\eta^{15}$, $\eta^{26}$, and $\eta^{37}$.
In addition, one must specify the relative shape of the three two-planes which is parameterized by the complex-structure moduli
\begin{equation}
u_1=\frac{r_5}{r_1}\,,
\qquad
u_2=\frac{r_6}{r_2}\,,
\qquad
u_3=\frac{r_7}{r_3}\,.
\end{equation}
In terms of the K\"ahler and complex-structure moduli, the metric becomes
\begin{align}\label{6Dmetric}
ds_6^2
=\sum_{\substack{a=1\\ a\neq 4}}^{7} (e^a)^2
=
\sum_{i=1}^3t_i\Big(u_i^{-1}(\eta^i)^2+u_i(\eta^{i+4})^2\Big) \,, 
\end{align}
which is organized into three orthogonal two-planes.

At this point, one can already see the correspondence between the natural degrees of freedom in the two dual descriptions. In the massive theory, one has eight scalar degrees of freedom, namely the seven moduli $s^i$, which measure the deformations of the internal three-cycles, together with the dilaton. In the massless theory, these are mapped to the three K\"ahler moduli $t_i$, the three complex-structure moduli $u_i$, the radius $r_4$ of the untwisted $S^1$ direction, and the dilaton.

Considering this, the K\"ahler form \eqref{JReOmega} takes the following form in the twisted basis
\begin{equation}\label{Jeta}
J=t_3\eta^{37}-t_1\eta^{15}-t_2\eta^{26} \,,
\end{equation}
and upon taking the exterior derivative, one finds that only the component along $\eta^{15}$ is non-closed, while $d(\eta^{26})=d(\eta^{37})=0$.
It then follows from \eqref{dF2nonzero} that
\begin{equation}
dJ=-\frac{t_1}{m_0}\,d\widetilde{F}_2\,.
\end{equation}

Unlike $J$, $\Omega$ depends not only on the products of radii but also on their ratios
\begin{equation}\label{ReOmegaImomega}
\begin{split}
\mathrm{Re}\,\Omega
&=
\sqrt{\frac{t_1 t_2 t_3}{u_1 u_2 u_3}}
\left(
u_3\,\eta^{127}
+u_2\,\eta^{136}
-u_1\,\eta^{235}
-u_1u_2u_3\,\eta^{567}
\right)\,,\\[0.5em]
\mathrm{Im}\,\Omega
&=
\sqrt{\frac{t_1 t_2 t_3}{u_1 u_2 u_3}}
\left(
-\eta^{123}
-u_2u_3\,\eta^{167}
+u_1u_3\,\eta^{257}
+u_1u_2\,\eta^{356}
\right)\,.
\end{split}
\end{equation}
Finally, the compatibility conditions take the standard form
\begin{equation}\label{vol6}
J\wedge\Omega=0\,,
\qquad
\mathrm{vol}_6\equiv-\frac{i}{8}\,\Omega\wedge\overline{\Omega}
=-\frac{1}{3!}J^3
=t_1t_2t_3\,\eta^{123567}\,.
\end{equation}
The first relation states that $J$ is indeed of type $(1,1)$ with respect to the almost complex structure defined by $\Omega$ while the second fixes the normalization of $\Omega$ relative to $J$ and determines the six-dimensional orientation.


\subsubsection{Differential relations and torsion classes}

We now determine the intrinsic torsion of the dual geometry by computing the exterior derivatives of the SU(3) structure forms. It is more convenient to work in the orthonormal frame \eqref{orthonormalframe}, rather than in the left-invariant basis $\eta^i$, since in this frame the SU(3) structure is written directly in a moduli-dependent form.

Using the Maurer--Cartan equation in the orthonormal frame \eqref{deAppendix}, together with the non-vanishing metric fluxes in \eqref{metricfluxes2}, one finds
\begin{equation}
de^1=\widehat\tau^1{}_{27}\,e^{27}+\widehat\tau^1{}_{36}\,e^{36}\,,
\qquad
de^5=\widehat\tau^5{}_{67}\,e^{67}+\widehat\tau^5{}_{23}\,e^{23} \,,
\end{equation}
while the remaining one-forms are closed. The hatted coefficients are the metric-dressed structure constants introduced in \eqref{deAppendix}, and the non-vanishing ones have the following explicit form
\begin{equation}\label{dressedtaus}
\widehat\tau^1{}_{27}\equiv \frac{r_1}{r_2r_7}\tau^1{}_{27}\,,
\quad\,\,\,\,\,
\widehat\tau^1{}_{36}\equiv \frac{r_1}{r_3r_6}\tau^1{}_{36}\,,
\quad\,\,\,\,\,
\widehat\tau^5{}_{67}\equiv \frac{r_5}{r_6r_7}\tau^5{}_{67}\,,
\quad\,\,\,\,\,
\widehat\tau^5{}_{23}\equiv \frac{r_5}{r_2r_3}\tau^5{}_{23}\,.
\end{equation}

We next compute the exterior derivatives of the SU(3) structure forms, and from \eqref{JReOmega} it follows immediately that
\begin{equation}\label{dJgeneralhat}
dJ
=
-d(e^{15})
=
\widehat\tau^1{}_{27}\,e^{257}
+\widehat\tau^1{}_{36}\,e^{356}
+\widehat\tau^5{}_{67}\,e^{167}
+\widehat\tau^5{}_{23}\,e^{123}\,.
\end{equation}
Likewise, using \eqref{eImOmega} one finds
\begin{equation}\label{dImdRehat}
d(\Re\Omega)
=
\left(
\widehat\tau^1{}_{27}
+\widehat\tau^1{}_{36}
-\widehat\tau^5{}_{67}
-\widehat\tau^5{}_{23}
\right)e^{2367}\,,
\qquad
d(\Im\Omega)=0\,,
\end{equation}
and therefore $d\Omega=d(\Re\Omega)$.

We now interpret these differential relations in terms of the torsion classes of the SU(3) structure.
The intrinsic torsion is encoded in the failure of $J$ and $\Omega$ to be closed, and decomposes into five irreducible $\SU(3)$ representations, the torsion classes $W_1,\dots,W_5$ \cite{Gray:1980,Chiossi:2002tw}. In the orientation adopted here, which aligns with \cite{Lust:2004ig}, the SU(3) structure equations have the form
\begin{align}
dJ &= -\frac{3}{2}\,\Im\!\left(W_1\overline{\Omega}\right)+W_4\wedge J+W_3 \,,\label{dJgeneral}\\
d\Omega &= W_1\,J\wedge J + W_2\wedge J + \overline{W}_5\wedge\Omega \,, \label{dOmegageneral}
\end{align}
where $W_1$ is a complex scalar, $W_2$ is a primitive $(1,1)$-form, $W_3$ is a real primitive $(2,1)+(1,2)$ form, $W_4$ is a real one-form, and $W_5$ is a $(1,0)$-form.

To extract the scalar torsion class $W_1$ we wedge \eqref{dOmegageneral} with $J$.
Using the primitivity of the $W_2$ which implies $W_2\wedge J^2=0$, together with $J\wedge\Omega=0$, one finds
\begin{equation}
d\Omega\wedge J = W_1J^3 \,,
\end{equation}
and using the volume relation \eqref{vol6} this gives
\begin{equation}\label{W1}
W_1=\frac{\Delta}{6}\,,
\qquad
\Delta=\widehat\tau^1{}_{27}
+\widehat\tau^1{}_{36}
-\widehat\tau^5{}_{67}
-\widehat\tau^5{}_{23}\,.
\end{equation}
Next, the term $\overline{W}_5\wedge\Omega$ in \eqref{dOmegageneral} captures the $(3,1)$ component of $d\Omega$, while the explicit computation in \eqref{dImdRehat} shows that $d\Omega$ is a real four-form of pure $(2,2)$ type.
Its $(3,1)$ component therefore vanishes and hence $W_5=0$.
Substituting this together with \eqref{W1} back into \eqref{dOmegageneral} and comparing with the explicit expression for $d\Omega$, one reads off
\begin{equation}\label{W2}
W_2=\frac{\Delta}{3}\left(2e^{15}-e^{26}+e^{37}\right) \,.
\end{equation}
Finally, wedging \eqref{dJgeneral} with $J$, and using $J\wedge\Omega=0$ together with the primitivity of $W_3$, which implies $J\wedge W_3=0$, gives
\begin{equation}
dJ\wedge J = W_4\wedge J^2 \,.
\end{equation}
In our case, the explicit expression for $dJ$ in \eqref{dJgeneralhat} implies that
\begin{equation}\label{JdJ}
dJ\wedge J=0 \,,
\end{equation}
since every term vanishes upon wedging with $J$ and thus it follows that $W_4\wedge J^2=0$.
Writing $W_4$ as one-form expansion, one can see that since the five-forms appearing in $W_4\wedge J^2$ are linearly independent, all coefficients of $W_4$ must vanish, thus $W_4=0$.
Considering that the structure equation \eqref{dJgeneral} reduces to
\begin{equation}
W_3=dJ-\frac32 W_1\Im\Omega \,,
\end{equation}
where we used that $W_1$ is real so $\text{Im}(W_1\Omega)=-W_1\text{Im}\,\Omega$, and the explicit expression is 
\begin{equation}\label{W3}
\begin{split}
W_3
&=
\left(\widehat{\tau}^1{}_{27}-\frac{\Delta}{4}\right)e^{257}
+\left(\widehat{\tau}^1{}_{36}-\frac{\Delta}{4}\right)e^{356}
+\left(\widehat{\tau}^5{}_{67}+\frac{\Delta}{4}\right)e^{167}
+\left(\widehat{\tau}^5{}_{23}+\frac{\Delta}{4}\right)e^{123} \,.
\end{split}
\end{equation}
We therefore conclude that, for the SU(3) structure inherited from the G$_2$ splitting, the intrinsic torsion is characterized by
\begin{equation}
W_1\neq 0\,,
\qquad
W_2\neq 0\,,
\qquad
W_3\neq 0\,,
\qquad
W_4=W_5=0\,.
\end{equation}
More specifically, the structure defines a half-flat $\mathrm{SU}(3)$ structure, since the following closure conditions are satisfied \cite{Chiossi:2002tw},
\begin{equation}
d(J\wedge J)=0 \,,
\qquad
d\,\mathrm{Im}\,\Omega=0 \,,
\end{equation}
 up to a constant phase redefinition of $\Omega$ as one may choose a phase convention in which $\text{Re}\,\Omega$ and $\text{Im}\,\Omega$ are interchanged.

Having determined the torsion classes, we can now compute the scalar curvature of the six-dimensional twisted space $X_6$ \cite{BEDULLI20071125} which is\footnote{One can explicitly compute the norms of the torsion classes,
\[
|W_1|^2=\frac{\Delta^2}{6^2}\,,\qquad
|W_2|^2=\frac{2\Delta^2}{3}\,,\qquad
|W_3|^2=(\widehat\tau^1{}_{27})^2+(\widehat\tau^1{}_{36})^2+(\widehat\tau^5{}_{67})^2+(\widehat\tau^5{}_{23})^2-\frac{\Delta^2}{4}\,.
\]
Substituting these expressions into the curvature formula, one finds that all $\Delta$ terms cancel precisely.}
\begin{equation}
\begin{split}
    R_6&=\frac{15}{2}\vert W_1\vert^2-\frac{1}{2}\vert W_2\vert^2-\frac{1}{2}\vert W_3\vert^2 \\
    &=-\frac{1}{2}\left((\widehat\tau^1{}_{27})^2+(\widehat\tau^1{}_{36})^2+(\widehat\tau^5{}_{67})^2+(\widehat\tau^5{}_{23})^2\right)  \,.
\end{split}
\end{equation}
The result reproduces the expression obtained independently in \eqref{eq:R7explicit}.


\paragraph{G$_2$ perspective.}
Using the SU(3) structure determined in subsection \ref{G2splitting}, we can evaluate the exterior derivatives of the G$_2$ forms in \eqref{PsiJOmega} and rewrite them in terms of $J$ and $\Omega$.
One finds
\begin{equation}
d\Phi=-\,e^4\wedge dJ+d(\Re\Omega)\,,
\qquad
d\Psi=J\wedge dJ+e^4\wedge d(\Im\Omega)=0\,,
\end{equation}
where the vanishing of the second expression follows from \eqref{dImdRehat} together with \eqref{JdJ}, and using \eqref{dJgeneralhat} we further obtain
\begin{equation}
d\Phi
=
\widehat\tau^1{}_{27}e^{2457}
+\widehat\tau^1{}_{36}e^{3456}
+\widehat\tau^5{}_{67}e^{1467}
+\widehat\tau^5{}_{23}e^{1234}
+\Delta\,e^{2367} \,.
\end{equation}
We now compute the G$_2$ torsion classes following the conventions of \cite{bryant2003g2} from the closure relations of G$_2$ structures
\begin{equation}
d\Phi=\widetilde{W}_0\star_7\Phi+3\widetilde{W}_1\wedge\Phi+\star_7\widetilde{W}_3 \,,
\qquad
d\star_7\Phi=4\widetilde{W}_1\wedge\star_7\Phi+\star_7\widetilde{W}_2 \,,
\end{equation}
where the torsion classes $\widetilde{W}_i$ are respectively a $0$-, $1$-, $2$-, and $3$-form transforming in the $\mathbf{1}$, $\mathbf{7}$, $\mathbf{14}$, and $\mathbf{27}$ representations of G$_2$ \cite{FernandezGray1982}.
Since $d\star_7\Phi=d\Psi=0$ it follows immediately that $\widetilde{W}_1=\widetilde{W}_2=0$.
To determine $\widetilde{W}_0$, one wedges the first equation with $\Phi$ and uses that $\Phi\wedge\star_7\widetilde{W}_3=0$ by definition of the $\mathbf{27}$ representation.
In this way one finds the non-vanishing torsion classes
\begin{equation}
\widetilde{W}_0=\frac{2}{7}\Delta \,,
\qquad
\star_7\widetilde{W}_3=d\Phi-\widetilde{W}_0\Psi \,,
\end{equation}
and the presence of only these torsion classes shows that the G$_2$ structure is co-calibrated\footnote{Interestingly, related co-calibrated G$_2$ and half-flat SU(3) structures also appear in the study of M-theory and massive type IIA compactifications on twisted orbifolds to four dimensions, where both reductions give rise to $\mathcal{N}=1$ STU-models and admit a non-trivial overlap in their effective descriptions \cite{Danielsson:2014ria}.
The relation between half-flat manifolds and G$_2$ spaces has also been discussed in \cite{Chiossi:2004ig,Ali:2006gd}.}.
From this, one can compute the curvature of the induced G$_2$ structure,
\begin{equation}
    R_7=\frac{21}{8}\widetilde{W}_0^2-\frac{1}{2}|\widetilde{W}_3|^2
    =R_6 \,,
\end{equation}
which agrees with the six-dimensional curvature since the seven-dimensional metric is locally a direct product of the twisted space $X_6$ with the untwisted $S^1$ direction parameterized by $e^4$.


\subsection{Superpotential in massless IIA via T-dualities}\label{modulistabdual}

Having established the geometry of the dual background, we can now use it to reconstruct the superpotential in the massless type IIA theory.
The most efficient way to proceed is to start from the known superpotential $P$ in the massive IIA compactification, write it explicitly in terms of the moduli, and then perform the double T-duality to determine the superpotential $\widetilde P$ in the massless dual frame.

In the present setup, all quantities are expressed in ten-dimensional Einstein frame.
For this reason, the T-duality rules are slightly more involved than in string frame, and we have summarized the necessary transformations in Appendix~\ref{AppendixDualities}.
For the specific double T-duality \eqref{eq:Tduality15}, we apply the rules presented in subsection~\ref{AppendixMatching}.
The radii along the dualized directions transform according to \eqref{eq:MetricTdualityEinsteinFrame} as
\begin{equation}\label{Tdr1}
   r'_1=e^{-3\phi/8}r_1^{-3/4}r_5^{1/4} \,,
   \qquad
   r'_5=e^{-3\phi/8}r_5^{-3/4}r_1^{1/4} \,,
\end{equation}
while for the undualized internal directions \eqref{eq:UndualizedMetricTdualityEinsteinFrame} gives
\begin{equation}\label{Tdr2}
   r'_n
   =
   e^{\phi/8}(r_1r_5)^{1/4}r_n\,,
   \qquad n\neq 1,5 \,.\,.
\end{equation}
The dilaton and the internal volume transform as
\begin{equation}\label{Tdr3}
   e^{\phi'}
   =
   \frac{e^{\phi/2}}{r_1 r_5} \,,
   \qquad
   \mathrm{vol}(X_7)^{\prime}
   =
   \prod_{i=1}^7 r_i^{\prime}
   =
   e^{-\phi/8}(r_1r_5)^{-1/4}\mathrm{vol}(X_7)\,.
\end{equation}
In principle, one could first rewrite the superpotential \eqref{eq:SuperpotentialExplictSi} entirely in terms of the radii, as was done for the scalar potential. However, since the superpotential is naturally expressed in terms of the three-cycle moduli $s^i$, and since their dependence on the radii is already known from \eqref{eq:stor}, it is more convenient to dualize the moduli $s^i$ directly. In this way, we obtain the massless type IIA superpotential written in terms of the radii:
\begin{equation}
   P\left(s'_i,\phi'\right)
   \quad\longrightarrow\quad
   \widetilde{P}\left(r_i,\phi\right)\,,
   \qquad i=1,\dots,7 \,.
\end{equation}
Using \eqref{eq:stor} together with the Einstein frame T-duality rules one finds
\begin{align}\label{eq:sdual}
\begin{alignedat}{3}
s_1'&=e^{-\phi/8}r_1^{-1/4}r_5^{3/4}r_2r_7 \,,\qquad
&s_2'&=e^{3\phi/8}(r_1r_5)^{3/4}r_3r_4r_7 \,,\qquad
&s_3'&=e^{-\phi/8}r_1^{3/4}r_5^{-1/4}r_6r_7 \,,\\
s_4'&=e^{-\phi/8}r_1^{-1/4}r_5^{3/4}r_3r_6 \,,\qquad
&s_5'&=e^{-\phi/8}r_1^{3/4}r_5^{-1/4}r_2r_3 \,,\qquad
&s_6'&=e^{-5\phi/8}r_1^{-1/4}r_5^{-1/4}r_4 \,,\\
s_7'&=e^{3\phi/8}(r_1r_5)^{3/4}r_2r_4r_6 \,,
\end{alignedat}
\end{align}
and substituting these expressions into the massive type IIA superpotential in \eqref{eq:SuperpotentialExplictSi} we find the massless type IIA superpotential written in components
\begin{align}\label{eq:SuperpotentialMapped2}
\widetilde{P}
=
\frac{1}{4\mathrm{vol}(X_7)^2}
\Bigg[
&\mathrm{vol}(X_7)\left(
h_1\frac{r_1}{r_2r_7}
+h_3\frac{r_5}{r_6r_7}
+h_4\frac{r_1}{r_3r_6}
+h_5\frac{r_5}{r_2r_3}
\right)\nonumber
\\[4pt]
&+e^{\phi/4}\left(
f_4^1\,r_2r_5r_7
+f_4^3\,r_1r_6r_7
+f_4^4\,r_3r_5r_6
+f_4^5\,r_1r_2r_3
\right)
\\[4pt]
&+e^{3\phi/4}\left(
m_0\,r_2r_3r_4r_6r_7
+f_4^2\,r_1r_3r_4r_5r_7
+f_4^7\,r_1r_2r_4r_5r_6
\right)
+e^{-\phi/4}f_4^6\,r_4
\Bigg] \,.\nonumber
\end{align}


\subsection{Moduli stabilization and families of solutions}\label{modulistabilization2}

We now analyze the dual supersymmetric AdS vacua by extremizing $\widetilde{P}$ with respect to the dilaton and the seven radii,
\begin{equation}
    \partial_{\phi}\widetilde{P}=0\,,
    \qquad
    \partial_{r^i}\widetilde{P}=0\,,
    \qquad
    i=1,\dots,7\,.
\end{equation}
Solving these equations, one finds that the string coupling is
\begin{equation}\label{dilatonscaling2}
     g_s\equiv e^{\phi}=\frac{2^{5/4}h}{\sqrt3}\left(\frac{m_0M}{N^3R^3}\right)^{1/4} \,,
\end{equation}
and unlike the massive IIA frame in \eqref{dilatonscaling}, the dual coupling contains an unconstrained flux in the numerator, which makes it possible to find parametrically strongly coupled regimes.

Next, after performing the Weyl rescaling \eqref{WeylRescaling}, the stabilized string frame radii are
\begin{align}\label{stringradii2}
    r_{S,1}&=r_{S,5}=\left(\frac{m_0M}{2NR}\right)^{1/4} \,,
    \qquad
    r_{S,2}=r_{S,6}=\left(\frac{2MN}{m_0R}\right)^{1/4}\,,
    \qquad
    r_{S,4}=\frac{2^{7/4}G}{(m_0MNR)^{1/4}}\,, \nonumber \\
    r_{S,3}&=\left(1+\sqrt{3}\right)\left(\frac{MR}{2m_0N}\right)^{1/4} \,,
    \qquad
    r_{S,7}=\left(-1+\sqrt{3}\right)\left(\frac{MR}{2m_0N}\right)^{1/4} \,,
\end{align}
while the vacuum expectation value is the same as in the massive vacua in \eqref{vacuumexpectationvalue}
\begin{equation}\label{vacuumexpectationvalue2}
    \langle \widetilde{V}\rangle=-4\widetilde{P}^2
    =
    -\frac{h^6m_0^4}{27}\frac{1}{(GMNR)^2} \,.
\end{equation}
Considering the Kaluza--Klein masses associated with the different internal radii, and retaining only the parametric dependence, we obtain
\begin{align}
    \frac{\langle \widetilde{V}\rangle}{m_{\text{KK},\{1,5\}}^2}
    \sim \frac{1}{NR}\,, \quad\,\,\,\,
    \frac{\langle \widetilde{V}\rangle}{m_{\text{KK},\{2,6\}}^2}
    \sim \frac{1}{R}\,, \quad\,\,\,\,
    \frac{\langle \widetilde{V}\rangle}{m_{\text{KK},4}^2}
    \sim \frac{G^2}{MNR}\,, \quad\,\,\,\,
    \frac{\langle \widetilde{V}\rangle}{m_{\text{KK},\{3,7\}}^2}
    \sim \frac{1}{N}\,.
\end{align}
These findings are also verified in Appendix~\ref{DualVacua} by performing the double T-duality on the massive vacuum of subsection~\ref{modulistabilization}.

\paragraph{Families of solutions.}
In what follows, we study the parametric regimes characterized by the behavior of the string coupling, the radii, and the scale-separation condition, and determine the corresponding conditions on the fluxes.

\begin{enumerate}
    \item \textbf{Classical solution 1.}

    In this regime, the solution is parametrically weakly coupled, scale separated, and all string frame radii become large
    \begin{equation}
        e^{\phi}\ll 1\,,
        \qquad
        r_{S,i}\gg 1\,,
        \qquad
        \frac{\langle \widetilde{V}\rangle}{m_{\text{KK},i}^2}\ll 1\,,
        \qquad \forall i \,,
    \end{equation}
    and these conditions are satisfied whenever
    \begin{equation}
    \sqrt{MNR}<G^2<MNR 
    \qquad \& \qquad 
    NR<M<N^3R^3 \,.
    \end{equation}

    \item \textbf{Classical solution 2.}

    In this regime, the solution is weakly coupled, scale separated, and all three-cycles are parametrically large, even though two of the radii become small
    \begin{equation}
        e^{\phi}\ll 1\,,
        \quad
        s_{S,i}\gg 1\,,
        \quad
        r_{S,\{1,5\}}\ll 1\,,
        \quad
        r_{S,\{2,3,4,6,7\}}\gg 1\,,
        \quad
        \frac{\langle \widetilde{V}\rangle}{m_{\text{KK},i}^{2}}\ll 1\,,
        \quad \forall i \,.
    \end{equation}
    This is realized for
    \begin{equation}
    \frac{(NR)^{3/2}}{M^{1/2}}<G^2<MNR
    \qquad \& \quad\,
    M<NR
    \qquad \& \quad\,
    R<MN
    \qquad \& \quad\,
    N\leq \sqrt{R}\,,
    \end{equation}
    while the last two conditions are not unique, since two alternative constraints can also be imposed.
    In this solution, although two of the radii become parametrically small, the three-cycles remain large and scale separation is still satisfied.

    \item \textbf{Strong coupling, large radii, and scale separation.}

    In this regime we require parametrically strong coupling, all string frame radii large, and scale separation
    \begin{equation}
        e^{\phi}\gg 1\,,
        \qquad
        r_{S,i}\gg 1\,,
        \qquad
        \frac{\langle \widetilde{V}\rangle}{m_{\text{KK},i}^{2}}\ll 1\,,
        \quad \forall i \,,
    \end{equation}
    and this is achieved when the fluxes satisfy
    \begin{equation}\label{eq:G2window}
    \sqrt{MNR}<G^2<MNR 
    \qquad \& \qquad
    M>N^3R^3 \,.
    \end{equation}
    This is the regime we were looking for, and it provides the appropriate setting for a possible uplift to eleven-dimensional supergravity, which is explored in \cite{GTRINGAS}.
\end{enumerate}


\subsubsection{Superpotential in differential-form language}

From the superpotential in \eqref{eq:SuperpotentialMapped2}, one can already identify the flux sectors of the massless theory.
The first line contains the metric-flux contributions associated with the geometric sector of the dual background, and hence with the curvature contribution to the scalar potential in \eqref{eq:R7explicit}. The second line reproduces the $\widetilde{F}_4$ sector, the third line gives the $\widetilde{F}_2$ contribution, and the final term is identified with the $\widetilde{F}_6$ sector\footnote{The superpotential \eqref{eq:SuperpotentialMapped2} should also reproduce the scalar potential obtained by direct dimensional reduction in \eqref{eq:potentialmasslessIIAexplicit}. We will not analyze this matching in detail here, since Appendix~\ref{AppendixMatching} already shows that the duality maps the scalar potential of the massive theory to that of the massless theory, and the same conclusion is expected to hold for the corresponding superpotentials. Nevertheless, it is straightforward to check that, at the vacuum, the dilaton dependence in \eqref{eq:SuperpotentialMapped2} reproduces the scaling appearing in \eqref{eq:potentialmasslessIIAexplicit}.}.

We now decompose the superpotential into its different sectors
\begin{equation}
    \widetilde{P}
    =
    \frac{1}{4\mathrm{vol}(X_7)^2}
    \big[
    \widetilde{P}_{W}
    +e^{3\phi/4}\widetilde{P}_{2}
    +e^{\phi/4}\widetilde{P}_{4}
    +e^{-\phi/4}\widetilde{P}_{6}
    \big] \,,
\end{equation}
and reconstruct each of them in differential-form language.
The first contribution is associated with the $\mathrm{SU}(3)$-structure; its dependence on the flux and radii is analogous to that of the curvature contribution to the massless scalar potential in \eqref{eq:R7explicit}, indicating that it captures the torsional sector of the dual background.
From the differential-form point of view, the relevant object is $dJ$ in \eqref{dJgeneralhat}, which is determined by the same metric-flux appearing in the first line of \eqref{eq:SuperpotentialMapped2}.
Moreover, once the overall volume is written explicitly, the remaining radius dependence is precisely the one encoded in $\mathrm{Re}\,\Omega$ in \eqref{ReOmegaImomega}. Therefore, the geometric part of the superpotential can be written in differential-form language, and a direct computation gives
\begin{equation}
\widetilde{P}_{W}
=
\int_{X_7} e^4\wedge \mathrm{Re}\,\Omega \wedge dJ
=
\mathrm{vol}(X_7)\left(
h_1\frac{r_1}{r_2r_7}
+h_3\frac{r_5}{r_6r_7}
+h_4\frac{r_1}{r_3r_6}
+h_5\frac{r_5}{r_2r_3}
\right)\,.
\end{equation}
In deriving this expression, we also use the normalization of the invariant top form
\begin{equation}
\int_{X_7}\eta^{1234567}=1\,.
\end{equation}

We next turn to the RR flux contributions. For the four-form sector, one finds
\begin{equation}
\widetilde{P}_{4}
=
\int_{X_7}\mathrm{Im}\,\Omega\wedge \widetilde F_4
=
f_4^1\,r_2r_5r_7
+f_4^3\,r_1r_6r_7
+f_4^4\,r_3r_5r_6
+f_4^5\,r_1r_2r_3 \,.
\end{equation}
Similarly, using the expression for $\widetilde F_6$ in \eqref{F6def}, together with the vielbein along the untwisted direction $e^4=r_4\eta^4$, one obtains
\begin{equation}
\widetilde{P}_{6}
=
-\int_{X_7} e^4 \wedge \widetilde{F}_6
=
f_4^6\,r_4 \,.
\end{equation}
For the two-form sector, one may also write the contribution in differential-form language. Although $\widetilde F_2$ is not proportional to the K\"ahler form $J$, both are expanded on the same invariant two-form basis, and only the $\eta^{15}$ component is non-closed. As a result, the corresponding term in the superpotential is expressed as
\begin{equation}
\widetilde{P}_{2}
=
\int_{X_7} e^4\wedge \frac{1}{2}J\wedge J\wedge \widetilde{F}_2
=
m_0\,r_2r_3r_4r_6r_7
+f_4^2\,r_1r_3r_4r_5r_7
+f_4^7\,r_1r_2r_4r_5r_6 \,.
\end{equation}

Collecting all these terms and including the dilaton factors and the overall volume, the dual superpotential \eqref{eq:SuperpotentialMapped2}, within the $\mathrm{SU}(3)$-structure conventions adopted above, can be rewritten in differential-form language as
\begin{equation}\label{eq:SuperpotentialSU3form}
\begin{split}
\widetilde P
=
\frac{1}{4\mathrm{vol}(X_7)^2}
\int_{X_7}\Big(
& e^4\wedge \mathrm{Re}\,\Omega \wedge dJ 
+e^{\phi/4}\mathrm{Im}\,\Omega\wedge \widetilde F_4\\
&+e^{3\phi/4}e^4\wedge\frac12\,J\wedge J\wedge \widetilde F_2
-e^{-\phi/4}e^4\wedge \widetilde F_6
\Big)\,.
\end{split}
\end{equation}

Let us now comment on the structure of this superpotential\footnote{This superpotential should also follow from the bispinor formalism of \cite{VanHemelryck:2022ynr}, upon rewriting the bispinors in SU(3) structure language following \cite{Dibitetto:2018ftj}.}.
In a standard six-dimensional $\mathrm{SU}(3)$ structure compactification, the RR four-form contribution to the superpotential is expected to arise through a term of the form $\int_{X_6} J\wedge F_4$, since both $J$ and $F_4$ are internal forms on $X_6$, see\footnote{It is expected to have a form $W_{\mathrm{SU}(3)}\sim
\int_{X_6}\Omega\wedge\bigl(H+i\,dJ\bigr)
+\int_{X_6}e^J\wedge F_{\mathrm{RR}}$, where the second integral is a polyform, so upon expanding it one obtains terms of the form $e^J\wedge F_{\mathrm{RR}}\sim J\wedge J\wedge F_2+J\wedge F_4+F_6$.} for example \cite{Grimm:2004ua}.
In the present setup however, the dual flux $\widetilde F_4$ is not supported entirely on the six-dimensional space, but instead carries one leg along the untwisted direction, $\widetilde F_4 \sim e^4 \wedge \Xi_3$, as shown in \eqref{F4dual}.
For this reason, the combination $e^4 \wedge J \wedge \widetilde F_4$ vanishes identically, and the relevant pairing instead involves the three-form sector on $X_6$ giving in our case the term $\int_{X_7}\mathrm{Im}\Omega\wedge\widetilde F_4$.


\section{Summary and outlook}\label{section5}

In this work, we constructed new parametrically scale-separated AdS$_3$ vacua in type IIA supergravity and analyzed their properties.
Our main goal was to construct massless type IIA AdS$_3$ vacua arising from G$_2$ compactifications, with stabilized moduli and parametric scale separation, containing only O6-planes and no non-geometric fluxes, thus making a possible M-theory uplift geometrically more transparent.

Our starting point was a new massive type IIA compactification on a G$_2$ holonomy toroidal orbifold with four smeared O6-planes, for which we reviewed and explained why previously studied orbifolds are not suitable for our purposes.
Within this setup, we analyzed the resulting three-dimensional effective theory, showed how tadpole cancellation is implemented, and identified parametrically classical families of vacua with stabilized moduli and scale separation.

After performing double T-duality, these massive type IIA solutions map to massless type IIA backgrounds containing only O6-planes, metric fluxes, and RR fields $F_2$, $F_4$, and $F_6$.
In the dual frame, we found that the geometry is locally described by a six-dimensional twisted space with half-flat $\mathrm{SU}(3)$ structure, more precisely of Iwasawa type, together with an untwisted circle direction, while globally it is quotiented by a non-trivial $\widehat{\mathbb{Z}}_2$ action inherited from the original orbifold.
We identified the invariant form subspaces under the global involution and used them to construct the relevant SU(3) structure data.
It is interesting that the six-dimensional subspace in the massless background has an algebra isomorphic to the one arising in the scale-separated AdS$_4$ uplift to M-theory in \cite{Cribiori:2021djm}.
Unlike in that case, the setup also contains an $F_4$ flux, which is the only flux threading the untwisted circle and is therefore responsible for stabilizing the corresponding radius.

Next, starting from the integrated superpotential in the massive theory, we derived the corresponding integrated superpotential in massless type IIA by performing T-dualities, and rewrote it in differential-form language using the geometry of the dual background.
The main result of this work, obtained by minimizing the dual superpotential, is a family of solutions in the massless theory that is scale-separated, has large radii, and is parametrically strongly coupled, thereby allowing for an uplift to eleven-dimensional supergravity.

There are several interesting directions in which the present work could be extended.
Although we are currently studying the M-theory uplift in \cite{GTRINGAS}, it would also be interesting to analyze the corresponding eight-dimensional geometry in more detail, which is expected to be related in some way to a Spin(7) structure.

A further direction would be to explore other double or quadruple T-dualities.
Although such dualities may lead to backgrounds with a net O4-plane contribution, it would be interesting to determine whether there exist configurations in which moduli stabilization and scale separation can still be realized, and whether the resulting dual background remains geometric.


\section*{Acknowledgments}
I would like to thank Fotis Farakos and Thomas Van Riet for discussions on topics related to this paper, and Vincent Van Hemelryck for pointing out important aspects of the dual solution that I had overlooked.
I am especially grateful to Timm Wrase for the valuable in person discussions during the preparation of this paper.
This work is supported in part by NSF Grant PHY-2210271 and by the Lehigh University CORE Grant (Grant ID: COREAWD40).


\appendix


\section{\texorpdfstring{Dimensional reduction of IIA bosonic action}{}}\label{app:AppendixII}
In this part of the appendix we perform the dimensional reduction of type IIA supergravity with O6-planes in order to obtain the three-dimensional scalar potential in both the massive and massless cases.

We start from the bosonic type IIA action in string frame
\begin{align}\label{SFrameAction}
    S&=\frac{1}{2\kappa_{10}^2}\int \text{d}^{10}X\sqrt{-G^S}e^{-2\phi}\left(R_{10}+4(\partial\phi)^2
    - \frac{1}{2}\vert H_3\vert^2
    - \frac{1}{4}e^{2\phi}\sum_{q}\vert F_q\vert^2
    \right)\,,
\end{align}
with $2\kappa^2_{10}=(2\pi)^7\alpha^{\prime 4}$.
We work in conventions in which the RR field strengths $F_q$ in type IIA are taken to be purely internal magnetic fluxes, so that
\begin{align}
    \text{IIA} \quad &:\quad q=0,2,4,6 \,.
\end{align}
For the O6-planes, the DBI and Wess--Zumino actions take the form
\begin{align}\label{sources}
    S_{\text{loc}}&=-\mu_{\text{O}6}\int\text{d}^{7}\xi\, e^{-\phi}\sqrt{-P[G^S]}+\mu_{\text{O}6}\int C_{7}\,,
\end{align}
where $P[G^S]$ denotes the pull-back of the ten-dimensional string frame metric onto the worldvolume, $\xi$ are the worldvolume coordinates, and $\mu_{\text{O}6}=-2\mu_{\text{D}6}<0$ is the tension and charge of the O6-plane source. For the Dp-brane tension we have $\mu_{\text{D}p}=(2\pi)^{-p}\left(\alpha^{\prime}\right)^{-\frac{p+1}{2}}$.

Next, in our conventions we first transform the ten-dimensional action from string frame to Einstein frame and then perform the dimensional reduction. To do so, we use the Weyl rescaling
\begin{equation}\label{WeylRescaling}
    G_{MN}^S=e^{\phi/2}G_{MN}^E \,.
\end{equation}
Ignoring the Wess--Zumino term and rewriting the action in the ten-dimensional Einstein frame, the bulk action takes the form
\begin{equation}\label{EinsteinFrameAction}
\begin{split}
    S_{E}&=\frac{1}{2\kappa_{10}^2}\int \text{d}^{10}X\sqrt{-G}\Bigg(R_{10}-\frac{1}{2}(\partial\phi)^2
    - \frac{1}{2}e^{-\phi}\vert H_3\vert^2
    - \frac{1}{4}e^{\frac{5-q}{2}\phi}\sum_{q}\vert F_q\vert^2 \\
    &\hspace{2.1cm}
    -2\kappa_{10}^2\mu_{\text{O}6}e^{\frac{p-3}{4}\phi}\sum_i\frac{1}{\sqrt{g_{3,i}}} \Bigg)
    \,.
\end{split}
\end{equation}
The DBI contribution of the O6-planes has been rewritten as a bulk term by smearing the localized sources over the internal space, so that the worldvolume integrals are replaced by ten-dimensional integrals with the appropriate volume factors. The sum over $i$ in the O6-plane term denotes the different O6-planes, or equivalently the different internal three-dimensional subspaces transverse to each O6-plane.

To perform the dimensional reduction using the metric ansatz in \eqref{dimrednastz}.
The resulting three-dimensional effective action takes the form
\begin{equation}
    S_3=4\pi\int d^3x\sqrt{-g_3}\left(\frac{1}{2}R_3-G_{IJ}\partial_{\mu}\varphi^I\partial^{\mu}\varphi^J-V_E\right)\,,
\end{equation}
where we have set $2\pi\sqrt{\alpha^{\prime}}=1$, and we will focus on the scalar potential. For the massive type IIA backgrounds considered here, only $H_3$, $F_0$, and $F_4$ are non-vanishing. Moreover, in the smeared approximation the underlying G$_2$ orbifold metric is Ricci-flat, so that $R_7=0$. The scalar potential then becomes
\begin{equation}\label{EffectiveactionmassiveIIA}
\begin{split}
    V_E&=\frac{1}{\mathrm{vol}(X_7)^2}\Bigg(
    \frac{1}{4}e^{-\phi}\vert H_3\vert^2
    +\frac{1}{8}e^{\frac{5}{2}\phi}\vert F_0\vert^2
    +\frac{1}{8}e^{\frac{1}{2}\phi}\vert F_4\vert^2
    +\frac{\mu_{\text{O}6}}{4\pi}e^{\frac{3}{4}\phi}\sum_i\frac{1}{\sqrt{g_{3,i}}}\Bigg)\,.
\end{split}
\end{equation}
Decomposing the scalar potential as
\begin{equation}
    V_E=V_{H_3}+V_{F_0}+V_{F_4}+V_{\text{O}6}\,,
\end{equation}
the individual contributions are given explicitly by
\begin{equation}\label{eq:potentialmassiveIIAexplicit}
\begin{split}
    V_{H_3}&=\frac{e^{-\phi}h^2}{4\mathrm{vol}(X_7)^2}\left(\frac{1}{s_1^{2}}+\frac{1}{s_3^{2}}+\frac{1}{s_4^{2}}+\frac{1}{s_5^{2}} \right) \,, \\
    V_{F_0}&=\frac{e^{\frac{5}{2}\phi}m_0^2}{8\mathrm{vol}(X_7)^2} \,, \\
    V_{F_4}
    &=\frac{e^{\frac{1}{2}\phi}}{8\mathrm{vol}(X_7)^4}\left( G^2\left(s_1^2+s_3^2+s_4^2+s_5^2\right)
    +N^2s_2^2
    +M^2s_6^2
    +R^2s_7^2\right) \,, \\
    V_{O6}
    &=\frac{\mu_{\text{O}6}}{4\pi}\frac{e^{\frac{3}{4}\phi}}{\mathrm{vol}(X_7)^2}\left(\frac{1}{s_1}+\frac{1}{s_3}+\frac{1}{s_4}+\frac{1}{s_5}\right) \,.
\end{split}
\end{equation}


\subsection{Curvature and 3D scalar potential from massless IIA}

In this subsection we compute the dimensionally reduced scalar potential of the massless theory. We begin by evaluating the curvature of the internal space, which is non-vanishing in this case due to the presence of metric fluxes.

To compute the internal Ricci scalar, it is convenient to pass to an orthonormal frame $e^a=r_a\eta^a$, where $\eta^a$ are the twisted one-forms, so that
\begin{equation}
    ds_7^2=\sum_a (e^a)^2 \,.
\end{equation}
The geometry is then encoded in the non-closure of the one-forms,
\begin{equation}\label{deAppendix}
\begin{split}
de^a
=r_a d\eta^a
=\frac12r_a\,\tau^a{}_{bc}\,\eta^b\wedge\eta^c
=\frac12\tau^a{}_{bc}\frac{r_a}{r_br_c}\,e^b\wedge e^c
=\frac12\widehat\tau^a{}_{bc}\,e^b\wedge e^c \,,
\end{split}
\end{equation}
where we used that $dr_a=0$ and $\eta^a=e^a/r_a$. The hatted structure constants are metric-dressed structure constants.
We now introduce the Levi--Civita spin connection through the Cartan equation
\begin{equation}
de^a+\omega^a{}_b\wedge e^b=0,
\qquad
\omega_{ab}=-\omega_{ba}\,,\quad\quad
\omega^a{}_b=\omega^a{}_{bc}\,e^c \,,
\end{equation}
and substituting \eqref{deAppendix} into it, one finds
\begin{equation}\label{omegaabc}
\omega_{abc}=\frac12\left(\widehat\tau_{abc}+\widehat\tau_{bca}-\widehat\tau_{cab}\right),
\end{equation}
where indices are lowered with the flat metric $\delta_{ab}$. The curvature two-form is then
\begin{equation}
R^a{}_b=d\omega^a{}_b+\omega^a{}_c\wedge\omega^c{}_b \, .
\end{equation}
Taking the exterior derivative of $\omega^a{}_b$, and using that the spin-connection coefficients $\omega^a{}_{bc}$ are constant in the orthonormal frame, we obtain
\begin{equation}
R^a{}_b=\frac{1}{2}\left(\omega^a{}_{bc}\widehat\tau^c{}_{ij}
+\omega^a{}_{ci}\omega^c{}_{bj}
-\omega^a{}_{cj}\omega^c{}_{bi}\right)e^i\wedge e^j
=\frac{1}{2}R^a{}_{bij}e^i\wedge e^j \, .
\end{equation}
Contracting gives the Ricci tensor,
\begin{equation}
R_{bd}=R^a{}_{bad}
=
\omega^a{}_{bc}\widehat\tau^c{}_{ad}
+\omega^a{}_{ca}\omega^c{}_{bd}
-\omega^a{}_{cd}\omega^c{}_{ba}.
\end{equation}
Substituting \eqref{omegaabc} into this expression and simplifying gives the Ricci tensor in terms of the coefficients $\widehat\tau^a{}_{bc}$,
\begin{equation}
2R_{cd} = -\widehat\tau^b{}_{ac}\widehat\tau^a{}_{bd} 
-\delta^{bg}\delta_{ah}\widehat\tau^h{}_{gc}\widehat\tau^a{}_{bd}
+\frac{1}{2}\,\delta^{ah}\delta^{bj}\delta_{ci}\delta_{dg}\widehat\tau^i{}_{aj}\widehat\tau^g{}_{hb} \,,
\end{equation}
in agreement with \cite{Andriot:2020wpp}.
In the present case the twisted space is unimodular, so that $\widehat\tau^a{}_{ab}=0$.
As a result, the trace terms that appear in the general expression for the Ricci tensor vanish. Contracting once more with $\delta^{cd}$, the internal Ricci scalar reduces to
\begin{equation}
R=\delta^{cd}R_{cd}
=
-\frac{1}{4}\widehat\tau_{abc}\widehat\tau^{abc}-\frac{1}{2}\widehat\tau_{abc}\widehat\tau^{bac} \,.
\end{equation}
From \eqref{deAppendix}, and for the non-zero metric fluxes considered in our example in \eqref{metricfluxes2}, we have
\begin{equation}
\widehat\tau^1{}_{27}=\tau^1{}_{27}\frac{r_1}{r_2r_7}\,,\,\,\,\quad
\widehat\tau^1{}_{36}=\tau^1{}_{36}\frac{r_1}{r_3r_6}\,,\,\,\,\quad
\widehat\tau^5{}_{67}=\tau^5{}_{67}\frac{r_5}{r_6r_7}\,,\,\,\,\quad
\widehat\tau^5{}_{23}=\tau^5{}_{23}\frac{r_5}{r_2r_3} \,.
\end{equation}
Therefore, only the first term in the expression for $R_7$ contributes, and each non-zero structure constant enters quadratically. This gives
\begin{equation}\label{eq:R7explicit}
\begin{split}
    R\equiv R_7&=-\frac12\left[\left(\tau^1{}_{27}\frac{r_1}{r_2r_7}\right)^2
    +\left(\tau^1{}_{36}\frac{r_1}{r_3r_6}\right)^2
    +\left(\tau^5{}_{67}\frac{r_5}{r_6r_7}\right)^2
    +\left(\tau^5{}_{23}\frac{r_5}{r_2r_3}\right)^2\right] \,,
\end{split}
\end{equation}
and the metric fluxes can be simplified further using \eqref{metricfluxes2}.

Following the discussion in the main part of this Appendix, the scalar potential of the massless theory, now including the curvature contribution from dimensional reduction, takes the explicit form
\begin{equation}
\begin{split}
    V_{E}&=\frac{1}{\mathrm{vol}(X_7)^2}\Bigg(-\frac{1}{2}R_{7}
    +\frac{1}{8}e^{\frac{3}{2}\phi}\vert \tilde{F}_2\vert^2
    +\frac{1}{8}e^{\frac{1}{2}\phi}\vert \tilde{F}_4\vert^2
    +\frac{1}{8}e^{-\frac{1}{2}\phi}\vert \tilde{F}_6\vert^2
    +\frac{\mu_{\widetilde{\text{O}}6}}{4\pi}e^{\frac{3}{4}\phi}\sum_i\frac{1}{\sqrt{g_{3,i}}}\Bigg)\,,
\end{split}
\end{equation}
and we decompose it as
\begin{equation}
    V_{3}=V_{R_7}+V_{\tilde{F}_2}+V_{\tilde{F}_4}+V_{\tilde{F}_6}+V_{\widetilde{\text{O}}6} \,.
\end{equation}
Collecting all contributions to the scalar potential we obtain
\begin{equation}\label{eq:potentialmasslessIIAexplicit}
\begin{split}
V_{R_7}
&=
-\frac{1}{2\mathrm{vol}(X_7)^2}
R_7,\\[0.4em]
V_{\tilde{F}_2}
&=
\frac{e^{\frac32\phi}}{8\mathrm{vol}(X_7)^2}
\left(
\frac{m_0^2}{r_1^2r_5^2}
+\frac{N^2}{r_2^2r_6^2}
+\frac{R^2}{r_3^2r_7^2}
\right),\\[0.4em]
V_{\tilde{F}_4}
&=
\frac{e^{\frac12\phi}}{8\mathrm{vol}(X_7)^2}\frac{G^2}{r_4^2}
\left(
\frac{1}{r_1^2r_2^2r_3^2}
+\frac{1}{r_1^2r_6^2r_7^2}
+\frac{1}{r_2^2r_5^2r_7^2}
+\frac{1}{r_3^2r_5^2r_6^2}
\right),\\[0.4em]
V_{\tilde{F}_6}
&=
\frac{e^{-\frac12\phi}}{8\mathrm{vol}(X_7)^2}
\frac{M^2}{r_1^2r_2^2r_3^2r_5^2r_6^2r_7^2},\\[0.4em]
V_{\widetilde{\mathrm{O}}6}
&=
\frac{\mu_{\widetilde{\mathrm{O}}6}}{4\pi}
\frac{e^{\frac34\phi}}{\mathrm{vol}(X_7)^2}
\left(
\frac{1}{r_1r_2r_3}
+\frac{1}{r_1r_6r_7}
+\frac{1}{r_2r_5r_7}
+\frac{1}{r_3r_5r_6}
\right).
\end{split}
\end{equation}


\section{T-duality in the Einstein frame}\label{AppendixDualities}

\subsection{Fluxes}

In order to implement the T-dualities of the fluxes and keep track of the signs, it is convenient to write a generic $p$-form as
\begin{equation}\label{Findicesnotation}
    F_p = \frac{1}{p!}F_{\mathbf{p},m_1\dots m_p}dy^{m_1\dots m_p} \,.
\end{equation}
According to this basis, and the flux ansatz in \eqref{F4nonminimal0}, the nonvanishing flux components in our configuration for the NSNS flux  is
\begin{equation}\label{Hdual1}
    H_{127}= h \,,\quad\quad 
    H_{567}= -h \,,\quad\quad
    H_{136}= h \,,\quad\quad
    H_{235}= -h \,,
\end{equation}
and similarly for the four-form RR flux
\begin{equation}
\begin{split}
F_{\textbf{4},3456}&= +f_4^1dy^{3456} \,,\quad\quad
    F_{\textbf{4},1256}= -f_4^2dy^{1256} \,,\quad\quad
    F_{\textbf{4},1234}= -f_4^3dy^{1234} \,,\\ 
    F_{\textbf{4},2457}&= +f_4^4dy^{2457} \,, \quad\quad
    F_{\textbf{4},1467}= -f_4^5dy^{1467} \,, \quad\quad
    F_{\textbf{4},2367}= +f_4^6dy^{2367} \,,\\
    F_{\textbf{4},1357}&= +f_4^7dy^{1357} \,. 
\end{split}
\end{equation}
whose specific values for our model are given in \eqref{fluxx}.

T-duality along a direction $y^i$ acts on an RR $p$-form as
\begin{equation}\label{RR_Tduality_operator}
T_i: \,\, F_p \quad\mapsto\quad \widetilde{F}_p=\left(\iota_i+dy^i\wedge\right)F=\widetilde{F}_{p-1}+\widetilde{F}_{p+1} \, ,
\end{equation}
where $\iota_i$ denotes contraction along $\partial_{y^i}$.
In components, $\iota_i$ removes an index $i$ from a form, producing a $(p-1)$-form, while
$dy^i\wedge$ adds an index $i$, producing a $(p+1)$-form. 
Thus, for our case we apply the T-duality map along $y^1$ and $y^5$, in the order $y^5$ followed by $y^1$. This choice fixes the overall sign convention in the RR sector, and the remaining RR fluxes map as follows
\begin{align}\label{RRmapping}
\begin{split}
    F_{\textbf{4},3456}& \quad\mapsto\quad \widetilde{F}_{\textbf{4},1346}=+f_4^1\eta^{1346}\\
    F_{\textbf{4},1256}& \quad\mapsto\quad \widetilde{F}_{\textbf{2},26}=-f_4^2\eta^{26}\\
    F_{\textbf{4},1234}& \quad\mapsto\quad \widetilde{F}_{\textbf{4},2345}=-f_4^3\eta^{2345}\\
    F_{\textbf{4},2457}& \quad\mapsto\quad \widetilde{F}_{\textbf{4},1247}=+f_4^4\eta^{1247}\\
    F_{\textbf{4},1467}& \quad\mapsto\quad \widetilde{F}_{\textbf{4},4567}=-f_4^5\eta^{4567}\\
    F_{\textbf{4},2367}& \quad\mapsto\quad \widetilde{F}_{\textbf{6},123567}=+f_4^6\eta^{123567}\\
    F_{\textbf{4},1357}& \quad\mapsto\quad
    \widetilde{F}_{\textbf{2},37}=+f_4^7\eta^{37}
\end{split}
\end{align}
We note that the flux signs remain unchanged.


\subsection{Metric and dilaton}\label{metricanddilaton}

In the following discussion, we focus on the metric--dilaton sector relevant for our analysis. 
We denote the string frame metric by $G^{S}$ and the Einstein frame metric by $G^{E}$, while primed fields indicate that T-duality has been applied.

\paragraph{String frame.} We apply the standard Buscher rules in string frame following \cite{Wecht:2007wu}. Under a T-duality along an internal direction $y_m$ we have the following transformations
\begin{equation}\label{eq:TdualityStringFrame}
   G^{S\prime}_{mm}=\frac{1}{G^{S}_{mm}} \,,
   \qquad
   G^{S\prime}_{nn}=G^{S}_{nn} \,,
   \qquad
   e^{\phi'}=\frac{e^{\phi}}{\sqrt{G^{S}_{mm}}} \,,
\end{equation}
and $y_n$ denotes any other internal direction that is not dualized.
More generally, under $k$-number of T-dualities along internal directions $y_i$, the dilaton transforms as
\begin{equation}\label{TdualityDilatonStringframe}
   e^{\phi'}
   =
   e^{\phi}\prod_{i=1}^k \frac{1}{\sqrt{G^{S}_{y_i y_i}}}
   =
   \frac{e^{\phi}}{\sqrt{\prod_{i=1}^k G^{S}_{y_i y_i}}} \,.
\end{equation}

\paragraph{Dilaton in Einstein frame.} In our analysis we work in ten-dimensional Einstein frame rather than string frame, so these relations cannot be applied directly to the dimensionally reduced scalar potential and superpotential.
Since the string frame and Einstein frame metrics are related by a Weyl rescaling, the T-duality transformation of the Einstein frame metric acquires additional dilaton-dependent factors.

After T-duality, the dual Einstein frame metric is obtained from the dual string frame metric through
\begin{equation}\label{WeylDualityEinsteinString}
   G^{E\prime}_{MN}=e^{-\phi'/2}G^{S\prime}_{MN} \,.
\end{equation}
We rewrite the Buscher transformation of the ten-dimensional dilaton in terms of Einstein frame metric components; starting from \eqref{TdualityDilatonStringframe} and using the Weyl rescaling from string frame to Einstein frame, \eqref{WeylRescaling}, one finds
\begin{equation}
   \prod_{i=1}^k G^{S}_{y_i y_i}
   =
   e^{k\phi/2}\prod_{i=1}^k G^{E}_{y_i y_i}.
\end{equation}
Substituting this into \eqref{TdualityDilatonStringframe}, the dilaton transforms in Einstein frame as
\begin{equation}\label{eq:DilatonTdualityEinsteinFrame}
   e^{\phi'}
   =
   e^{\frac{4-k}{4}\phi}
   \left(\prod_{i=1}^k G^{E}_{y_i y_i}\right)^{-1/2}.
\end{equation}

\paragraph{Dualized directions in Einstein frame.}

Now, we explore how a direction in the Einstein frame transforms under T-duality. 
After duality, the Einstein frame and string frame metrics are related by \eqref{WeylDualityEinsteinString}, therefore, for one dualized direction $y_i$,
\begin{equation}
   G^{E\prime}_{y_i y_i}
   =
   e^{-\phi'/2}\,G^{S\prime}_{y_i y_i}
   =
   e^{-\phi'/2}\,\frac{1}{G^{S}_{y_i y_i}}.
\end{equation}
Using \eqref{web}, to replace $G^{S}_{y_i y_i}$ in the previous expression, and together with the expression for $e^{-\phi'/2}$ from \eqref{eq:DilatonTdualityEinsteinFrame}, one finds
\begin{equation}\label{eq:MetricTdualityEinsteinFrame}
\begin{split}
   G^{E\prime}_{y_i y_i}
   &=
   e^{\frac{k-8}{8}\phi}
   \left(\prod_{j=1}^k G^{E}_{y_j y_j}\right)^{1/4}
   \bigl(G^{E}_{y_i y_i}\bigr)^{-1} \\
   &=
   e^{\frac{k-8}{8}\phi}
   \bigl(G^{E}_{y_i y_i}\bigr)^{-3/4}
   \prod_{j\neq i}\bigl(G^{E}_{y_j y_j}\bigr)^{1/4} \,.
\end{split}
\end{equation}

\paragraph{Undualized directions in Einstein frame.} 
It is important to note that, although the undualized internal directions are invariant in string frame, cf.~\eqref{eq:TdualityStringFrame}, they are not invariant in Einstein frame. Indeed, after duality the Einstein frame metric is related to the dual string frame metric through \eqref{WeylDualityEinsteinString}, and therefore
\begin{equation}
   G^{E\prime}_{nn}
   =
   e^{-\phi'/2}G^{S\prime}_{nn}
   =
   e^{-\phi'/2}G^{S}_{nn} \,.
\end{equation}
Using the Weyl rescaling \eqref{WeylRescaling}, one finds
\begin{equation}
   G^{E\prime}_{nn}
   =
   e^{(\phi-\phi')/2}G^{E}_{nn}.
\end{equation}
Thus, even though the string frame component $G^{S}_{nn}$ is invariant, the Einstein frame component $G^{E}_{nn}$ acquires a non-trivial rescaling because the dilaton itself transforms under T-duality. More generally, for $k$ T-dualities along the internal directions $y_i$, using \eqref{eq:DilatonTdualityEinsteinFrame} one obtains
\begin{equation}\label{eq:UndualizedMetricTdualityEinsteinFrame}
   G^{E\prime}_{nn}
   =
   e^{k\phi/8}
   \left(\prod_{i=1}^k G^{E}_{y_i y_i}\right)^{1/4}
   G^{E}_{nn},
   \qquad n\notin y_i \,.
\end{equation}
Hence, in Einstein frame the undualized internal directions are rescaled by a common dilaton-dependent factor, even though they are left invariant in string frame.


\subsection{Matching of the 3D scalar potentials}\label{AppendixMatching}

A useful consistency check of the duality relations derived above is to verify that the three-dimensional scalar potential transforms correctly when passing from the massive type IIA frame to the massless one. More precisely, we would like to show that the scalar potential obtained in massive IIA, \eqref{eq:potentialmassiveIIAexplicit}, is mapped to the corresponding potential in massless IIA, \eqref{eq:potentialmasslessIIAexplicit}.

The simplest term to analyze is the NSNS contribution. In this case, the relevant part of the map is schematically
\begin{equation}
   V_{H_3}\left(r'_i,\phi'\right)
   \quad\longrightarrow\quad
   \widetilde{V}_R\left(r_i,\phi\right)\,,
   \qquad i=1,\dots,7 \,.
\end{equation}
That is, we expect the NSNS potential in the massive theory to be mapped to the internal-curvature contribution in the massless theory. The goal of this subsection is therefore to verify explicitly that the NSNS term in \eqref{eq:potentialmassiveIIAexplicit} reproduces the curvature contribution appearing in \eqref{eq:potentialmasslessIIAexplicit}.

For the specific double T-duality \eqref{eq:Tduality15}, one has $k=2$. The Einstein frame metric transformation for the dualized directions, cf.~\eqref{eq:MetricTdualityEinsteinFrame}, then gives
\begin{equation}
   G^{E\prime}_{11}
   =
   e^{-3\phi/4}
   \bigl(G^{E}_{11}\bigr)^{-3/4}
   \bigl(G^{E}_{55}\bigr)^{1/4} \,,
   \qquad
   G^{E\prime}_{55}
   =
   e^{-3\phi/4}
   \bigl(G^{E}_{55}\bigr)^{-3/4}
   \bigl(G^{E}_{11}\bigr)^{1/4} \,,
\end{equation}
while for the undualized internal directions, \eqref{eq:UndualizedMetricTdualityEinsteinFrame} gives
\begin{equation}
   G^{E\prime}_{nn}
   =
   e^{\phi/4}
   \bigl(G^{E}_{11}G^{E}_{55}\bigr)^{1/4}
   G^{E}_{nn},
   \qquad n\neq 1,5.
\end{equation}
Using the parametrization of the internal metric components
\begin{equation}
   G^{E}_{ii}=r_i^2,
\end{equation}
one obtains the explicit transformation laws presented in \eqref{Tdr1},\eqref{Tdr2} and \eqref{Tdr3}.
Using that the NSNS potential can be rewritten in terms of the radii of the massless frame as
\begin{align}
    V_{H_3}
    &=\frac{e^{-\phi^{\prime}}}{4\left(r_1^{\prime}r_2^{\prime}r_3^{\prime}r_4^{\prime}r_5^{\prime}r_6^{\prime}r_7^{\prime}\right)^2}
    \Bigg(\left(\frac{h^1}{r_1^{\prime}r_2^{\prime}r_7^{\prime}}\right)^{2}
    +\left(\frac{h^3}{r_5^{\prime}r_6^{\prime}r_7^{\prime}}\right)^{2}
    +\left(\frac{h^4}{r_1^{\prime}r_3^{\prime}r_6^{\prime}}\right)^{2}
    +\left(\frac{h^5}{r_2^{\prime}r_3^{\prime}r_5^{\prime}}\right)^{2} \Bigg) \nonumber \\
    &=\frac{e^{-\phi/2}r_1 r_5}{4\left(e^{-3\phi/8}r_1^{-3/4}r_5^{1/4}\times e^{-3\phi/8}r_5^{-3/4}r_1^{1/4}\times\left(e^{\phi/8}(r_1r_5)^{1/4}\right)^5r_2r_3r_4r_6r_7\right)^2} \nonumber\\
    &\times\Bigg(\left(\frac{h^1}{e^{-3\phi/8}r_1^{-3/4}r_5^{1/4}\left(e^{\phi/8}(r_1r_5)^{1/4}\right)^2r_2r_7}\right)^{2}
    +\left(\frac{h^3}{e^{-3\phi/8}r_5^{-3/4}r_1^{1/4}\left(e^{\phi/8}(r_1r_5)^{1/4}\right)^2r_6r_7}\right)^{2} \nonumber\\
    &
    +\left(\frac{h^4}{e^{-3\phi/8}r_1^{-3/4}r_5^{1/4}\left(e^{\phi/8}(r_1r_5)^{1/4}\right)^2r_3r_6}\right)^{2}
    +\left(\frac{h^5}{r_2r_3\left(e^{\phi/8}(r_1r_5)^{1/4}\right)^2e^{-3\phi/8}r_5^{-3/4}r_1^{1/4}}\right)^{2} \Bigg) \nonumber\\
    &=\frac{1}{4\mathrm{vol}(X_7)^2}
    \Bigg(\left(\frac{h^1r_1}{r_2r_7}\right)^2
    +\left(\frac{h^3r_5}{r_6r_7}\right)^2
    +\left(\frac{h^4r_1}{r_3r_6}\right)^2
    +\left(\frac{h^5r_5}{r_2r_3}\right)^2 \Bigg) \,.
\end{align}
This reproduces the curvature contribution of the massless theory in \eqref{eq:potentialmasslessIIAexplicit} using \eqref{eq:R7explicit}.

\subsection{Matching of massless IIA vacua}\label{DualVacua}

We now apply the double T-duality to the massive IIA vacua of subsection~\ref{modulistabilization} in order to determine the stabilized values of the moduli in the dual frame and verify the results of subsection~\ref{modulistabdual}.

Using the double T-duality in \eqref{TdualityDilatonStringframe}, the string coupling in \eqref{dilatonscaling} transforms as
\begin{equation}
e^{\phi'}=\frac{e^\phi}{r_{S,1}r_{S,5}}
=
\frac{2^{5/4}h}{\sqrt3}\left(\frac{m_0M}{N^3R^3}\right)^{1/4} \,.
\end{equation}
Next, applying the double T-duality \eqref{eq:TdualityStringFrame} to the stabilized string frame radii in \eqref{stringradii}, one finds that the dual radii and the total internal volume in string frame are
\begin{align}
    r'_{S,\{1,5\}}=\frac{1}{r_{S,\{1,5\}}}
    =\left(\frac{m_0M}{2NR}\right)^{1/4} \,,\qquad
    \mathrm{vol}'_S(X_7)=\frac{\mathrm{vol}_S(X_7)}{r_{S,1}^2r_{S,5}^2}
    =G\left(\frac{2^9M^5}{m_0^3N^3R^3}\right)^{1/4} \,,
\end{align}
and $r'_{S,i}=r_{S,i}$ for $i\neq 1,5$.
Then, the scale separation conditions become
\begin{equation}
\begin{split}
\frac{\langle V\rangle}{m_{\text{KK},\{i=1,5\}}^{\prime\,2}}
&\sim r_{S,i}^{-2}\,e^{-4\phi}\,\mathrm{vol}_S(X_7)^2\,\langle V\rangle
\sim \frac{1}{NR} \,, \\
\frac{\langle V\rangle}{m_{\text{KK},i}^{\prime\,2}}
&\sim \frac{\langle V\rangle}{m_{\text{KK},i}^{2}} \,,\qquad i\neq 1,5 \,.
\end{split}
\end{equation}
It is also straightforward to determine the string frame three-cycle moduli in the dual frame, since only $r_{S,1}$ and $r_{S,5}$ change
\begin{equation}
\begin{split}
s_{S,1}^{\prime}&=\frac{s_{S,_1}}{r_{S,1}^2}=s_{S,3}^{\prime}=\frac{s_{S,3}}{r_{S,5}^2}
=(\sqrt3-1)\left(\frac{M^3}{2m_0NR}\right)^{1/4} \,, \\
s_{S,4}^{\prime}&=\frac{s_{S,4}}{r_{S,1}^2}=s_{S,5}^{\prime}=\frac{s_{S,5}}{r_{S,5}^2}
=(1+\sqrt3)\left(\frac{M^3}{2m_0NR}\right)^{1/4} \,, \\
s_{S,2}^{\prime}&=s_{S,2}=4G\left(\frac{2MR}{m_0^3N^3}\right)^{1/4} \,,\\
s_{S,6}^{\prime}&=\frac{s_{S,6}}{r_{S,1}^2 r_{S,5}^2}
=2^{5/4}G\left(\frac{m_0M}{N^3R^3}\right)^{1/4} \,,\\
s_{S,7}^{\prime}&=s_{S,7}=4G\left(\frac{2MN}{m_0^3R^3}\right)^{1/4} \,.
\end{split}
\end{equation}


\bibliographystyle{JHEP}
\bibliography{refs}

\end{document}